\documentclass[superscriptaddress,amsmath,amssymb,aps,prl,twocolumn,floatfix,amsart]{revtex4-2}

\usepackage{graphicx}
\usepackage{float}
\usepackage{bm}
\usepackage[colorlinks, linkcolor=mycol, citecolor=mycol, urlcolor=mycol, breaklinks]{hyperref}
\usepackage{braket}
\usepackage{bbold}
\usepackage{xcolor} 
\usepackage{comment}
\usepackage{physics}
\usepackage{verbatim}
\usepackage{amsthm}
\newcommand{\mi}{ {\rm i} }
\newcommand{\me}{ {\rm e} }
\newcommand{\id}{\mathbb{1}}

\newtheorem{result}{Result}
\newtheorem{remark}{Remark}

\usepackage{dsfont}
\usepackage{orcidlink}

\usepackage{tikz}

\usepackage{amsmath}
\usepackage{amssymb}
\usepackage{ulem}

\definecolor{mycol}{RGB}{10,55,130}

\newcommand{\customsection}[1]{\textit{#1---}}

\begin{document}

\title{Detecting multilevel entanglement from light-based entanglement witnesses} 

\author{P.~Rosario~\orcidlink{0000-0002-7628-7373}}
\email{pedrorosario@estudante.ufscar.br}
\affiliation{Departamento de Física, Universidade Federal de S\~ao Carlos, Rodovia Washington Lu\'is, km 235—SP-310, 13565-905 S\~ao Carlos, SP, Brazil}
\affiliation{CESQ/ISIS (UMR 7006), CNRS and Universit\'{e} de Strasbourg, 67000 Strasbourg, France}

\author{R.~Bachelard~\orcidlink{0000-0002-6026-509X}}
\email{romain@ufscar.br}
\affiliation{Departamento de Física, Universidade Federal de S\~ao Carlos, Rodovia Washington Lu\'is, km 235—SP-310, 13565-905 S\~ao Carlos, SP, Brazil}

\date{\today}

\begin{abstract}
We introduce a set of electric-field based inequalities capable of detecting multilevel entanglement from a system of $N$ quantum emitters. We determine  that the polarization channel as well as the direction of detection can enhance entanglement detection,  a feature specific to multilevel systems. We demonstrate the efficiency of the witnesses to detect genuine multipartite entanglement by applying it to families of paradigmatic quantum states, such as Dicke states, singlet states and W-like states. The detection is not only robust to noise, but also applies to mixed entangled states. Our findings open up possibilities for the detection of entanglement without local measurements in systems of multilevel emitters such as superconducting qubits, Rydberg atoms or quantum dots.

\end{abstract}

\maketitle

\customsection{Introduction} Generating and detecting large-scale entanglement in many-body quantum systems is of fundamental interest. Recent experimental progress on multilevel structures~\cite{Anton_2018,Tian_2024,Alexey_2022,Forbes_2021,Can_2020,Chi_2022,Zhao_2024,Chang_2021,Monz_2022,Monz_2023,Heng_2025} has drawn a lot of attention on the potential multilevel entanglement for quantum information processing~\cite{Anton_2020,Wang_2020,Rey_2024,Gunhe_2018,Rey_2019,Eckner_2023}. Remarkably, multilevel atoms allow one to reduce circuit complexity~\cite{White_2009}, benefit from more robust quantum error correction~\cite{dan_2014} and enable the implementation of optimal quantum measurements~\cite{Martin_2022}. 

However, quantifying and detecting entanglement in high-dimensional setups is recognized as an NP-hard problem~\cite{Gurvits_2003}, since  for $N$ particles with $d$ level each, the size of the Hilbert space increases even faster than for two-level atoms, following the scaling $(\mathcal{H}^{d})^{\otimes N}$ with $d\geq 3$. As a consequence, entanglement witnesses have been investigated as a promising tool to detect multipartite entanglement without the need for state tomography. Formally, an entanglement witness is defined as follows: let $\mathcal{O}$ be an arbitrary collective operator, then $\mathcal{O}$ is an entanglement witness if a violation of the inequality $\text{tr}[\mathcal{O}\hat{\rho}]\geq 0$ corresponds to entanglement in $\hat{\rho}$~\cite{GUHNE_2009_1}. In this regard, for an ensemble of two-level systems, collective operators such as; structure factors~\cite{Macchiavello_2009}, collective spin \cite{Toth_TLS_2007}  and far-field electric field \cite{Rosario_2024} have been proposed as multipartite entanglement witnesses. Nevertheless, in the case of multilevel systems, due to their intrinsic complexity, few witnesses have been introduced to date. For example, collective operators of arbitrary spins have been used for the detection of entanglement, which has allowed to formulate  connections between multilevel structures and spin squeezing ~\cite{toth_multi_lvl_2011,toth_2014}.

\begin{figure}[t!]
\includegraphics[width=\columnwidth]{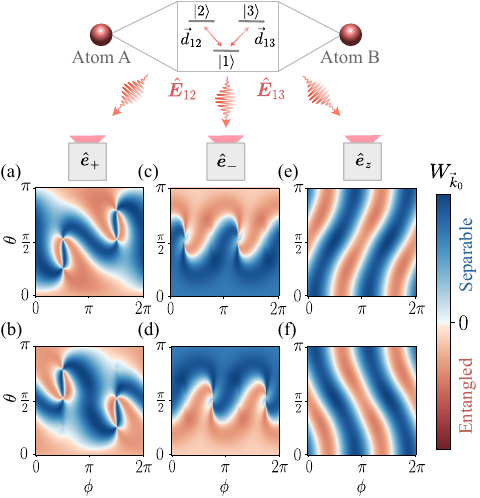}
    \caption{ Entanglement witness $W_{\vec{k}_{0}}$ for two three-level atoms prepared in the quantum state $\ket{\psi}=\frac{1}{\sqrt{3}}(\ket{11}+\ket{22}+\ket{13})$ with a relative distance between the emitters of $15/k_{0}$ along the $\hat{z}$ axis. The light emitted on both transitions is detected  along three different channels of light polarization: (a)-(b) right circular $\hat{{e}}_{+}$ [case (i)], (c)-(d) left circular $\hat{{e}}_{-}$  [case (ii)], and (e)-(f) linear $\hat{e}_{z}$ [case (iii)]. It is collected in the direction $\hat{{R}}=(\sin\theta\cos\phi,\sin\theta\sin\phi,\cos\theta)$, with $0\leq \theta\leq \pi$ and $0\leq \phi \leq 2\pi$, which is here stereographically mapped to a plane. See main text for details on the dipole orientations.}
    \label{fig:1}
\end{figure}

In this work, we introduce fundamental connections between  entanglement and the far field light emitted from a set of $N$ multilevel atoms. We show that the geometry between the detector and the atoms has a direct influence on the entanglement sensing. Thus, each direction of detection provides a different witness. Furthermore, contrary to two-level systems, we establish that light polarization plays a key role when witnessing entanglement
in multilevel systems. These results pave the way for efficient large-scale entanglement detection, where quantum tomography of the full-density matrix is no longer feasible. At the same time, our findings open the door for the understanding of light as a carrier of information related to non-classical quantum correlations.


\customsection{Witnesses for multipartite entanglement} Let us first present the theoretical framework for the derivation of our main result. A quantum emitter with $d$ levels and its Hilbert space $\mathcal{H}^{d}$ can be described using a variety of bases, ranging from Gell-Mann matrices to Weyl operators~\cite{Bertlmann_2008}. In our case, it is convenient to work with
{\it local orthogonal observables} (LOOs)~\cite{Sixia_2005}, a collection of $d^{2}$ Hermitian operators $\{\hat{G}_{m}\}_{m=1}^{d^{2}}$ with dimension $d\times d$ satisfying $\text{tr}\left[\hat{G}_{m}\hat{G}_{n}\right]=\delta_{mn}$ and $\sum_{m=1}^{d^{2}}\hat{G}_{m}\hat{G}_{m}=\id_d$. We denote the decomposition of a quantum state in this basis as~\cite{Sixia_2005,gunhe_2006_LOOs}
\begin{align}
\hat{\varrho}=\sum_{m=1}^{d^{2}}g_{m}\hat{G}_{m}.
\label{eq:local_def_1}
\end{align}
Using the canonical basis of the atomic states $\{\ket{\gamma}\}_{\gamma=1}^{d}$ and $1\leq \alpha < \beta \leq d$, LOOs can explicitly be chosen as:

\begin{align}
  \hat{G}_{m}=  \begin{cases}
        \frac{\left(\ketbra{\alpha}{\beta}+\ketbra{\beta}{\alpha}\right)}{\sqrt{2}}&\text{if}\ \  1\leq m \leq \frac{d(d-1)}{2}, \\
        \frac{\mi\left(\ketbra{\alpha}{\beta}-\ketbra{\beta}{\alpha}\right)}{\sqrt{2}} &\text{if}\ \  \frac{d(d-1)}{2} < m \leq d(d-1),\\
        \ketbra{\beta}{\beta} &\text{if}\ \  d(d-1) < m \leq d^{2}.
    \end{cases}
\end{align}

This set of LOOs actually corresponds to ladder and population operators of the atom, and this is at the basis of our main result: multipartite entanglement can be detected from the radiation from the collection of $N$-multilevel emitters. Indeed, the photon creation operator for a transition $\alpha \leftrightarrow \beta$ of a cloud of $N$-multilevel atoms is given by the (normalized) positive frequency part of the electric field in the far field~\cite{Agarwal1974,Bojer_2022}
\begin{align}
    \hat{\boldsymbol{E}}^{+}_{\alpha \beta}=(\hat{\boldsymbol{E}}^{-}_{\alpha \beta})^{\dagger}=(\hat{{R}}\times (\hat{{R}}\times \vec{d}_{\alpha \beta}))\sum_{\eta=1}^{N}\me^{-\mi \vec{k}_{0} \cdot \vec{r}_{\eta}}\hat{\Lambda}^{(\eta)}_{\alpha \beta},
    \label{eq:E_multi_lvl}
\end{align}
where $\hat{{R}}$ represents the unitary vector of the direction of detection, $\vec{r}_{\eta}$ the $\eta$-th atom position, $\vec{d}_{\alpha \beta}$ the dipolar moment for that transition, $\vec{k}_0=k_0\hat{{R}}$ the wavevector of the mode of observation ($k_0$ the transition wavenumber) and $\hat{\Lambda}_{\alpha \beta}=\ketbra{\alpha}{\beta}$ the interatomic ladder operator (with $\beta>\alpha$). Using  Eq.~\eqref{eq:E_multi_lvl}, we then introduce the electric field quadratures $\hat{X}_{\alpha \beta}$ and $\hat{Y}_{\alpha \beta}$ for the transition $\alpha \leftrightarrow \beta$:
\begin{align}
    &\hat{X}_{\alpha \beta}=\hat{{e}}_{\alpha\beta}\cdot\hat{\boldsymbol{E}}^{+}_{\alpha\beta}+(\hat{{e}}_{\alpha\beta}\cdot\hat{\boldsymbol{E}}^{+}_{\alpha \beta})^{\dagger},
    \label{eq:X}\\
    &\hat{Y}_{\alpha \beta}=\mi\hat{{e}}_{\alpha\beta}\cdot\boldsymbol{E}^{+}_{\alpha \beta}-\mi(\hat{{e}}_{\alpha\beta}\cdot\boldsymbol{E}^{+}_{\alpha\beta})^{\dagger},
    \label{eq:Y}
\end{align}
where $\hat{{e}}_{\alpha\beta}$ is a unitary polarization vector along which the light from the $\alpha \leftrightarrow \beta$ transition  is detected (such as circular, $\hat{{e}}_{\pm}$, linear, $\hat{{e}}_{z}$ or any combination of these).

After a proper normalization, Eqs.~\eqref{eq:X} and~\eqref{eq:Y} can be written as (see SM \cite{SM}):
\begin{align}
    &\hat{\boldsymbol{X}}_{\mu_{+}}=\frac{\hat{X}_{\alpha\beta}}{\sqrt{2}|\zeta_{\alpha\beta}|}=\sum_{\eta=1}^{N}\hat{\boldsymbol{G}}^{(\eta)}_{\mu_{+}},\\
&\hat{\boldsymbol{Y}}_{\mu_{-}}=\frac{\hat{Y}_{\alpha\beta}}{\sqrt{2}|\zeta_{\alpha\beta}|}=\sum_{\eta=1}^{N}\hat{\boldsymbol{G}}^{(\eta)}_{\mu_{-}}.
\end{align}
where we have introduced the phase-dependent LOOs: 

\begin{align}
    &\hat{\boldsymbol{G}}^{(\eta)}_{\mu_{+}}:= \frac{\me^{-\mi \vec{k}_{0}\cdot \vec{r}_{\eta}}\zeta_{\alpha \beta}\hat{\Lambda}^{(\eta)}_{\alpha \beta}+\me^{\mi \vec{k}_{0}\cdot \vec{r}_{\eta}}\zeta_{ \beta \alpha}\hat{\Lambda}^{(\eta)}_{\beta \alpha}}{\sqrt{2}|\zeta_{\alpha \beta}|}, \label{eq:G_1} \\
    &\hat{\boldsymbol{G}}^{(\eta)}_{\mu_{-}}:=\mi\frac{\me^{-\mi \vec{k}_{0}\cdot \vec{r}_{\eta}}\zeta_{\alpha \beta}\hat{\Lambda}^{(\eta)}_{\alpha \beta}-\me^{\mi \vec{k}_{0}\cdot \vec{r}_{\eta}}\zeta_{ \beta \alpha}\hat{\Lambda}^{(\eta)}_{\beta \alpha}}{\sqrt{2}|\zeta_{\alpha \beta}|}.
    \label{eq:G_2}
\end{align}
 The coefficients $1\leq \mu_{+}\leq d(d-1)/2$ and $d(d-1)/2 < \mu_{-}\leq d(d-1)$ stand for the mapping to a single index  $\mu:=\{\alpha\beta\}$, for simplicity. The complex factor $\zeta_{\alpha \beta}=(\zeta_{\beta\alpha })^{*}=\hat{e}_{\alpha\beta}\cdot(\hat{{R}}\times (\hat{{R}}\times \vec{d}_{\alpha \beta}))$ is here assumed to be non-zero, which occurs when  the vector $\hat{{e}}_{\alpha\beta}$ is non-orthogonal to the polarization of the radiated light on that transition. Indeed, due to the existence of selection rules and forbidden transitions, the quantity $\zeta_{\alpha \beta}$ is not defined for all pairs $(\alpha,\beta)$. 
Let $T_{1}$ be the number of atomic transition where $\vec{d}_{\alpha\beta}$ is physically defined and $T_{2}$ where it is not. Then, a complete set of LOOs, containing $d^{2}$ elements, is obtained by including $2T_{2}$ LOOs independent of the light polarization (i.e. $\hat{\boldsymbol{G}}^{(\eta)}_{\mu_{+}}$ and $\hat{\boldsymbol{G}}^{(\eta)}_{\mu_{-}}$ free of $\zeta_{\alpha\beta}\neq 0$) and $N$ population operators $\hat{\boldsymbol{G}}^{(\eta)}_{\mu_{z}}:=\hat{\Lambda}^{(\eta)}_{\beta \beta}$ [$d(d-1)<\mu_{z}\leq d^{2}$], so that $2T_{2}+2T_{1}+N=d^{2}$. Hence, the new family of LOOs is given by $\hat{\boldsymbol{G}}_{m}=$ $\{\hat{\boldsymbol{G}}^{(\eta)}_{\mu_{+}}, \hat{\boldsymbol{G}}^{(\eta)}_{\mu_{-}},\hat{\boldsymbol{G}}^{(\eta)}_{\mu_{z}}\}$. The set $\{\hat{\boldsymbol{G}}_{m}\}_{m=1}^{d^{2}}$ also satisfies the property $\text{tr}(\hat{\boldsymbol{G}}_{m}\hat{\boldsymbol{G}}_{n})=\delta_{mn}$, so they properly define a density matrix basis for Eq.~\eqref{eq:local_def_1} (see SM \cite{SM}). 
 
Using the previous developed theory, we are now ready to  formalize the connection between the light emitted from multilevel systems and multipartite entanglement:

\begin{result} 
Violation of the following inequality implies entanglement in $\hat{\rho}$ (see SM \cite{SM}):
\label{r:1}
\begin{align}
W_{\vec{k}_{0}}=\text{min}\{w_{1,\vec{k}_{0}}, w^{\bar{\boldsymbol{A}},\bar{\boldsymbol{B}},\bar{\boldsymbol{C}}}_{2,\vec{k}_{0}}, w^{\bar{\boldsymbol{A}},\bar{\boldsymbol{B}},\bar{\boldsymbol{C}}}_{3,\vec{k}_{0}}\}\geq 0,
    \label{eq:main_result}
\end{align}
{\it with } $W_{\vec{k}_{0}}$ {\it the minimum of the following observables}
\begin{align}
    &\nonumber w_{1,\vec{k}_{0}}\!=\!\sum_{\mu_{+}}(\Delta \hat{\boldsymbol{X}}_{\mu_{+}})^{2}\!+\!\sum_{\mu_{-}}(\Delta \hat{\boldsymbol{Y}}_{\mu_{-}})^{2}\!+\sum_{\mu_{z}}(\Delta \hat{\boldsymbol{Z}}_{\mu_{z}})^{2}\!\\
    &\nonumber-\! (d-1)N,\\
    &\nonumber w^{\bar{\boldsymbol{A}},\bar{\boldsymbol{B}},\bar{\boldsymbol{C}}}_{2,\vec{k}_{0}}\!=\!(N-1)\!\sum_{\mu_{a}}(\Delta \bar{\boldsymbol{A}}_{\mu_{a}})^{2}\! -\! \sum_{\mu_{b}}\langle \bar{\boldsymbol{B}}^{2}_{\mu_{b}}\rangle \!-\!\sum_{\mu_{c}}\langle \bar{\boldsymbol{C}}^{2}_{\mu_{c}}\rangle\! \\
    &\nonumber+\! N(N-1),\\
    &\nonumber w^{\bar{\boldsymbol{A}},\bar{\boldsymbol{B}},\bar{\boldsymbol{C}}}_{3,\vec{k}_{0}}\!=\!(N-1)\!\left(\sum_{\mu_{a}}(\Delta \bar{\boldsymbol{A}}_{\mu_{a}})^{2}\!+\!\sum_{\mu_{b}}(\Delta \bar{\boldsymbol{B}}_{\mu_{b}})^{2}\right) \\
    &\nonumber  -\!\sum_{\mu_{c}}\langle \bar{\boldsymbol{C}}^{2}_{\mu_{c}}\rangle +\! N(N-1).
\end{align}
\end{result}
Here, $(\Delta \hat{\boldsymbol{\mathcal{O}}})^{2}=\langle \hat{\boldsymbol{\mathcal{O}}}^{2}\rangle -\langle \hat{\boldsymbol{\mathcal{O}}}\rangle^{2}$ stands for the variance of $\hat{\boldsymbol{\mathcal{O}}}$, and the quantities $w^{\bar{\boldsymbol{A}},\bar{\boldsymbol{B}},\bar{\boldsymbol{C}}}_{2,\vec{k}_{0}}$ and $w^{\bar{\boldsymbol{A}},\bar{\boldsymbol{B}},\bar{\boldsymbol{C}}}_{3,\vec{k}_{0}}$ were defined using the modified second moment $\langle \bar{\boldsymbol{\mathcal{O}}}^{2}_{\mu_{k}}\rangle=\langle \hat{\boldsymbol{\mathcal{O}}}^{2}_{\mu_{k}}\rangle-\Big\langle  \sum_{\eta=1}^{N}\left(\hat{\boldsymbol{G}}^{(\eta)}_{\mu_{k}}\right)^{2}\Big\rangle$ and variance $(\Delta \bar{\boldsymbol{\mathcal{O}}}_{\mu_{k}})^{2}=(\Delta \hat{\boldsymbol{\mathcal{O}}}_{\mu_{k}})^{2}-\Big\langle  \sum_{\eta=1}^{N}\left(\hat{\boldsymbol{G}}^{(\eta)}_{\mu_{k}}\right)^{2}\Big\rangle$. We point out that Eq.~\eqref{eq:main_result} admits any permutation of the operators $\{\bar{\boldsymbol{A}}_{\mu_{a}},\bar{\boldsymbol{B}}_{\mu_{b}},\bar{\boldsymbol{C}}_{\mu_{c}}\}$, which are taken from the set $\{\bar{\boldsymbol{X}}_{\mu_{+}},\bar{\boldsymbol{Y}}_{\mu_{-}},\bar{\boldsymbol{Z}}_{\mu_{z}}\}$ with $a,b,c \in \{+,-,z\}$ (e.g. if $\bar{\boldsymbol{B}}=\bar{\boldsymbol{X}}$ then $\mu_{b}=\mu_{+}$). 

Two properties can be derived from Eq.~\eqref{eq:main_result}:

\begin{remark}
\label{remark:1}
     If $\hat{e}_{\alpha\beta},\vec{d}_{\alpha,\beta}\in \mathbb{R}$ , then $W_{\vec{k}_{0}}$ is independent of the measured light polarization $\hat{e}_{\alpha,\beta}$.
\end{remark} 
This condition is fulfilled when measuring  a linear polarization $\hat{{e}}_{z}$ (real vector) and its proof follows from the fact that in this scenario the factor that codifies the information related to the measured light polarization ($\zeta_{\alpha,\beta}$) is a real number, so it cancels out from Eqs.~\eqref{eq:G_1} and ~\eqref{eq:G_2}.

\begin{remark}
\label{remark:2}
 The witnesses $w_{1,\vec{k}_{0}}$, $w^{\bar{\boldsymbol{Z}},\bar{\boldsymbol{X}},\bar{\boldsymbol{Y}}}_{2,\vec{k}_{0}}$ and $w^{\bar{\boldsymbol{X}},\bar{\boldsymbol{Y}},\bar{\boldsymbol{Z}}}_{3,\vec{k}_{0}}$ are  always independent of the measured light polarization $\hat{e}_{\alpha\beta}$, i.e, even if $\zeta_{\alpha\beta}\notin \mathbb{R}$.
\end{remark}
In other words, they are  ``polarization-blind" and can detect entanglement independently of the polarization channel chosen for the detection (see SM for a proof~\cite{SM}).

Note that while the collective spin operators satisfy a set of inequalities similar to Eq.~\eqref{eq:main_result}~\cite{toth_multi_lvl_2011}, the present electric-field-based inequalities take into account the internal structure of the atom, which is determined by the dipole coupling. In particular, the optical phases and light polarization constitute new degrees of freedom which can be exploited to detect different families of entangled states, as we shall now discuss. 
Due to their diverse applications in quantum optics and quantum information, we consider both multipartite symmetric~\cite{Wei_2003,Popkov_2005,Raveh_2024} and asymmetric $d$-dimensional quantum states~\cite{Maurer_2001,Cabello_2002}.
\smallskip

\customsection{Symmetric (Dicke) states with white noise}
Let us first focus on the family of symmetric superradiant Dicke states, in the special case of a system with as many particles as there are individual levels, $N=d$. We consider a symmetric state with occupation number $(1,1,\hdots,1)$, where each single-particle level $\ket{1},\hdots,\ket{d}$ appears exactly once across the $N$ positions~\cite{Raveh_2024}. Such states are particularly useful for multiparty
quantum communication tasks and quantum metrology~\cite{Huangjun_2021}, and they have been realized experimentally for few particles~\cite{Hiesmayr_2016}. They write as $\ket{S_{N}}=\frac{1}{\sqrt{N!}}\sum_{\sigma \in \boldsymbol{S}_{N}}\ket{\sigma(1),\sigma(2),\hdots,\sigma(N)}\in (\mathcal{H}^{d})^{\otimes d}$,
with $\boldsymbol{S}_{N}$ the permutation symmetric group with $N!$ elements.
In addition, in order to simulate imperfections during the preparation of $\ket{S_{N}}$, we add a source of white noise, which maps the initial quantum state to a statistical mixture of density matrices: $\hat{\varrho}= p(\id_{d^{N}}/d^{N})+(1-p)\ketbra{S_{N}}{S_{N}}$.

Applying Result~\ref{r:1} to this symmetric state, we find that $p$ must satisfy the following condition for entanglement to be detected using witness~\eqref{eq:main_result} in direction $\vec{k}_{0}$, see SM~\cite{SM}:
\begin{align}
    p < \frac{(N-S_{\vec{k}_{0}})}{(N-1)+(N-S_{\vec{k}_{0}})}
    \label{eq:sym_p},
\end{align}
where $S_{\vec{k}_{0}}=\left|\sum_{\eta=1}^{N}\me^{\mi \vec{k}_{0}\cdot \vec{r}_{\eta}}\right|^{2}$ is the structure factor ($0 \leq S_{\vec{k}_{0}} \leq N^{2} $). Hence, for this state the witness is independent of the light polarization. Furthermore, the best direction to measure the witness is when $S_{\vec{k}_{0}}$ is minimum, which corresponds to the directions of destructive interference, $S_{\vec{k}_{0}}=0$. Note that although the average field is zero, a field is still detected due to quantum fluctuations (spontaneous emission).. For example, in 1D regular arrays with atoms at positions $\vec{r}_{\eta}=(\eta -1)a\hat{z}$, with $\eta=1,\hdots,N$ and $a$ the lattice step, one obtains $S_{\vec{k}_{0}}=|(1-\me^{\mi N k_{0z}a})/(1-\me^{\mi k_{0z}a})|^{2}$, which cancels for $N k_{0z}a=0 \textrm{ mod } 2\pi$. In this direction entanglement is detected provided that $p < N/(2N-1)$. In particular, for any $N$ multipartite entanglement is detected for $p\leq 1/2$, demonstrating the robustness of the witness.

\smallskip

\customsection{Antisymmetric (dark) states with white noise}
Let us now consider a family of antisymmetric states, specifically, $SU(d)$ singlet states. They are a useful resource for, e.g., the problems of $N$-strangers, secret sharing and liar detection~\cite{Cabello_2002}. Those states belong to the family of dark (subradiant) states and can be prepared recursively  using bipartite gates  or collective emission~\cite{Helmut_2017}, and are defined as $\ket{A_{N}}=\frac{1}{\sqrt{N!}}\sum_{\sigma \in \boldsymbol{S}_{N}}\text{sgn}(\sigma)\ket{\sigma(1),\sigma(2),\hdots,\sigma(N)}$. Similarly to the symmetric state $\ket{S_{N}}$, the local dimension of the qudits satisfies $d=N$ and the operation $\text{sgn}(\sigma)$ is equal to $1$ for cyclic permutations and $-1$ otherwise. As before, we include random white noise to $\ket{A_{N}}$, such that the new quantum state takes the form $\hat{\varrho}= p(\id_{d^{N}}/d^{N})+(1-p)\ketbra{A_{N}}{A_{N}}$.

Using Result~\ref{r:1}, the following condition is obtained for the detection of entanglement (see SM~\cite{SM})
\begin{align}
    p< \frac{(S_{\vec{k}_{0}}-N)}{(N-1)+(S_{\vec{k}_{0}}-N)}.
    \label{eq:asym_p}
\end{align}
Hence, differently from symmetric states, constructive interference are preferred directions for the detection of entanglement. For example, in 1D and 2D atomic configurations, the fully constructive interference condition, $S_{\vec{k}_{0}}=N^2$, is reached when considering an orthogonal observation, such that $\vec{k}_0\cdot (\vec{r}_{\eta_{1}}-\vec{r}_{\eta_{2}})=2\pi m,\ \forall \eta_{1},\eta_{2}=1,\hdots,N$ with $m$ any positive integer number. In this case, witness~\eqref{eq:main_result} is violated provided that $p< N/(N+1)$, a result in agreement with that obtained when using multilevel collective spin as entanglement witnesses~\cite{toth_multi_lvl_2011}. Remarkably, in the large $N$ limit, multipartite is detected whatever the level of noise, demonstrating once more the robustness of the light-based witness.

\smallskip

\customsection{$N$-qudit W-class states} W-states are among the most important quantum states in quantum information, since they present robustness against particle loss and are a key resource for many applications such as photonic error detection \cite{Vijayan_2020}, quantum memory \cite{Duan_2020}, quantum secret sharing \cite{Tsai_2019} and quantum communication. These states can be defined as~\cite{Kim_2008,Chuan_2020}
\begin{equation}
\ket{\text{W}_{d}}=\frac{1}{\sqrt{(d-1)N}}\sum_{i=1}^{N}\sum_{j=2}^{d}\ket{1}^{\otimes(i-1)}\otimes\ket{j}_{i}\otimes\ket{1}^{\otimes(N-i)},
\end{equation}
which belong to the Hilbert space $(\mathcal{H}^{d})^{\otimes N}$. Then,  and assuming $d\geq 3$, we find that the following condition must be satisfied for the inequality~\eqref{eq:main_result} to be violated (see SM \cite{SM}):

\begin{align}
    S_{\vec{k}_{0}}\leq\frac{N^{2}}{2N-1}.
\end{align}
Interestingly, this condition is independent of the internal dimension of the atom $d$. Furthermore, as for the symmetric Dicke states, the directions of destructive interference are best to detect the entanglement. 


\customsection{Polarization of light matters} While the previous states did not present polarization-dependent detection, let us now discuss a case where the detection is polarization sensitive. We consider the entangled quantum state of two particles $\ket{\psi}=\frac{1}{\sqrt{3}}(\ket{11}+\ket{22}+\ket{13})$, which belongs to the Hilbert space $\mathcal{H}^{3}\otimes\mathcal{H}^{3}$. The atoms are located along the $\hat{z}$ axis at the positions $\vec{r}_{1}=\vec{0}$ and $\vec{r}_{2}=15\vec{z}/k_{0}$ and possess a V internal structure, see illustration in Fig.~\ref{fig:1}. The transitions $\ket{1}\leftrightarrow \ket{2}$ and $\ket{1}\leftrightarrow \ket{3}$ are the only ones allowed, with orthogonal dipole moments $\vec{d}_{12}$ and $\vec{d}_{13}$ which we consider in two possible configurations for each measured polarization: (i) right-circular polarization $\hat{e}_{+}$, Fig.\ref{fig:1}(a) $\vec{d}_{12}=(\hat{y}+\hat{z})/\sqrt{2}$, $\vec{d}_{13}=(\hat{z}-\hat{y})/\sqrt{2}$ and Fig.\ref{fig:1}(b) $\vec{d}_{12}=(\hat{z}-\hat{y})/\sqrt{2}$, $\vec{d}_{13}=(\hat{y}+\hat{z})/\sqrt{2}$. (ii) left-circular polarization $\hat{e}_{-}$, Fig.\ref{fig:1}(c) $\vec{d}_{12}=(\hat{x}+\hat{y})/\sqrt{2}$ , $\vec{d}_{13}=(\hat{x}-\hat{y})/\sqrt{2}$ and Fig.\ref{fig:1}(d) $\vec{d}_{12}=(\hat{x}-\hat{y})/\sqrt{2}$, $\vec{d}_{13}=(\hat{x}+\hat{y})/\sqrt{2}$. (iii) linear polarization $\hat{e}_{z}$, Fig.\ref{fig:1}(e) $\vec{d}_{12}=\hat{e}_{-}$, $\vec{d}_{13}=\hat{e}_{z}$ and Fig.\ref{fig:1}(f) $\vec{d}_{12}=\hat{e}_{+}$, $\vec{d}_{13}=\hat{e}_{z}$.

The dependence of the witness on both the direction of observation $\hat{k}_0=k_0(\sin\theta\cos\phi, \sin\theta\sin\phi,\cos\theta)$ and on the polarization of the dipoles can be appreciated in Fig.\ref{fig:1}, where $W_{\vec{k}_{0}}$ is represented in the $(\theta,\phi)$ plane. 
Notably, we observe that not all directions $\hat{k}_0$ are suitable for entanglement detection.  
Note that the cases (i) and (ii) where real dipole vectors are considered generate a symmetric pattern by just interchanging the dipole orientations between the transitions and measuring the same channel of light polarization. In contrast, if both the dipole moment and the measured light polarization are complex vectors, then such a symmetry will be broken. In this case, a way to preserve the symmetry is by measuring a linear light polarization as illustrated in case (iii) of Fig.\ref{fig:1}. 

 In addition, we remark that when the dipole moment for each allowed transition is a real vector, measuring right circular light polarization from all allowed transitions will produce a mirror-like symmetry with left circular light polarization. For the quantum state in consideration, it can be seen by calculating \eqref{eq:main_result} as
 
\begin{align}
    \nonumber W^{\hat{{e}}_{\pm}} _{\vec{k}_{0}}&=-\frac{4}{3}\cos{(\vec{k}_{0}\cdot(\vec{r}_{1}+\vec{r}_{2})\mp 2\varphi_{12})}\\&+\frac{1}{9}\cos{(2\vec{k}_{0}\cdot \vec{r}_{2}\mp 2\varphi_{13})}+\frac{5}{9}.
    \label{eq:symmetry}
\end{align}
where $W^{\hat{{e}}_{\gamma}} _{\vec{k}_{0}}$ means that we are only measuring  $\gamma$ light  polarization from both transitions and $\varphi_{\alpha\beta}=\tan^{-1}(v_{2}/v_{1})$ is a phase induced by the dipole with $\{v_{1},v_{2}\}$ being the first and second components of the vector $\hat{{R}}\times (\hat{{R}}\times \vec{d}_{\alpha \beta})$ respectively. From Eq.\eqref{eq:symmetry} we note that the pattern generated by $\hat{e}_{+}$ is the same as $\hat{e}_{-}$ but submitted to a spatial point rotation  as presented in Fig.\ref{fig:2} considering $\vec{d}_{12}=(\hat{x}+\hat{z})/\sqrt{2}$ and $\vec{d}_{13}=(\hat{x}-\hat{z})/\sqrt{2}$. This result can be generalized to an arbitrary number of transitions and particles as follows:

\begin{figure}[t!]
\includegraphics[width=\columnwidth]{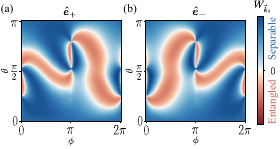}
    \caption{Two three-level atoms prepared in the quantum state $\ket{\psi}=\frac{1}{\sqrt{3}}(\ket{11}+\ket{22}+\ket{13})$ with a relative distance between the emitters of $15/k_{0}$ along the $\hat{z}$ axis. Following the protocol described in Fig.~\ref{fig:1}, the light is measured along the polarization channels (a) right circular and (b) left circular, which generates a symmetric mirror pattern for $W_{\vec{k}_{0}}$, as stated in Remark \ref{remark:3}. See main text for details on dipole configurations.}
    \label{fig:2}
\end{figure}

\begin{remark}
\label{remark:3}
    Let $W^{\hat{{e}}_{+}} _{\vec{k}_{0}}$ ($W^{\hat{{e}}_{-}} _{\vec{k}_{0}}$) denote Eq.\eqref{eq:main_result} when all the light emitted from the $T_{1}$ allowed transitions with $\vec{d}_{\alpha\beta}\in \mathbb{R}$ $\forall (\alpha,\beta)$ is measured along right-(left)-circular polarization. Then $W^{\hat{{e}}_{+}} _{\vec{k}_{0}}$ has  point mirror symmetry with $W^{\hat{{e}}_{-}} _{\vec{k}_{0}}$ in a $T_{1}$-dimensional space (see SM \cite{SM})
\end{remark}

A consequence of Remark \ref{remark:3} is that if we measure light composed only of the pair of polarizations $(\hat{{e}}_{+},\hat{{e}}_{z})$, for example, then the result will have mirror symmetry with  $(\hat{{e}}_{-},\hat{{e}}_{z})$, since $\hat{{e}}_{z}$ only introduces a real factor.

From the last observations, it follows that manipulating the dipole orientation and the channel of light polarization allows us to detect entanglement in different regions of space where the detector $\hat{R}$ can be located. This result introduces light polarization as a new degree of freedom to improve the detection of entanglement in multilevel systems. 

\customsection{Conclusion $\&$ perspectives}We have introduced a set of light-based entanglement witnesses for arbitrary multilevel systems. This result is a direct consequence of a direct mapping between local orthogonal observables and field quadrature operators. We demonstrate the efficiency of these witnesses by detecting the entanglement in the presence of quantum noise from a group of paradigmatic families of quantum states such as Dicke states, singlet states, and W-like states. For these states, the directions suitable for entanglement detection are related to the structure factor of the atomic ensemble. Furthermore, for certain states this detection relies on a judicious choice of the polarization channels in which the field is monitored -- an illustration of the importance of accounting for the internal structure of the emitters to detect multipartite entanglement. 

Our results open new perspectives for the study of large open quantum systems, where collective states can be generated, for example, via dipole-dipole interactions~\cite{Rey_2019,Rey_2024,Masson_2024}. In particular, the set of inequalities~\eqref{eq:main_result} could be transformed into multilevel spin squeezing parameters~\cite{toth_2014}, to quantify the potential gain for measurements of the probed quantum states~\cite{Nori_2010,L_pezze_2018,Vitagliano_2025}.

\begin{acknowledgments}
{\it Acknowledgments.---} P.R., and R.B. acknowledge the financial support of the São Paulo Research Foundation (FAPESP) (Grants No. 2022/12382-4, 2024/05564-4, 2025/00697-9, 2023/03300-7, 2022/00209-6). CAPES-COFECUB (CAPES, Grant No. 88887.711967/2022-00 and COFECUB, Grant No. Ph 997/23), and from the Brazilian CNPq (Conselho Nacional de Desenvolvimento Científico e Tecnológico), Grant No. 313632/2023-5.
\end{acknowledgments}
\bibliography{bib.bib}


\onecolumngrid
\newpage
\begin{center}
{\large{ {\bf Supplemental Material for: \\ Detecting multilevel entanglement from light based entanglement witnesses}}}

\setcounter{page}{1}
\vskip0.5\baselineskip{P. Rosario,$^{1,2}$ and R. Bachelard $^{1}$}

\vskip0.5\baselineskip{ {\it $^{1}$Departamento de Física, Universidade Federal de São Carlos,\\ Rodovia Washington Luís, km 235 - SP-310, 13565-905 São Carlos, SP, Brazil}}
\vskip0.1\baselineskip{{\it $^{2}$CESQ/ISIS (UMR 7006), CNRS and Universit\'{e} de Strasbourg, 67000 Strasbourg, France}}
\end{center}

\appendix
\section{Connecting far-field scattered electric field with LOOs}
Eq.\eqref{eq:X} and \eqref{eq:Y} in the main text, can explicitly be written as

\begin{align}
    &\hat{X}_{\alpha \beta}=\sum_{\eta=1}^{N}\left[e^{-\mi \vec{k}_{0}\cdot \vec{r}_{\eta}}\zeta_{\alpha \beta}\hat{\Lambda}^{(\eta)}_{\alpha \beta}+e^{\mi \vec{k}_{0}\cdot \vec{r}_{\eta}}\zeta_{ \beta \alpha}\hat{\Lambda}^{(\eta)}_{\beta \alpha}\right],
    \label{eq:X2}\\
    &\hat{Y}_{\alpha \beta}=\mi\sum_{\eta=1}^{N}\left[e^{-\mi \vec{k}_{0}\cdot \vec{r}_{\eta}}\zeta_{\alpha \beta}\hat{\Lambda}^{(\eta)}_{\alpha \beta}-e^{\mi \vec{k}_{0}\cdot \vec{r}_{\eta}}\zeta_{ \beta \alpha}\hat{\Lambda}^{(\eta)}_{\beta \alpha}\right],
    \label{eq:Y2}
\end{align}
 where  the complex factor $\zeta_{\alpha \beta}=\hat{e}_{\alpha\beta}\cdot(\hat{{R}}\times (\hat{{R}}\times \vec{d}_{\alpha \beta}))$ is a geometric factor associated with a specific light polarization and the detector position. From the previous expressions, we note that the argument inside the brackets (up to a normalization factor) can be used to define the LOOs:
\begin{align}
    &\hat{\boldsymbol{G}}^{(\eta)}_{\mu_{+}}:= \frac{\left(\me^{-\mi \vec{k}_{0}\cdot \vec{r}_{\eta}}\zeta_{\alpha \beta}\hat{\Lambda}^{(\eta)}_{\alpha \beta}+\me^{\mi \vec{k}_{0}\cdot \vec{r}_{\eta}}\zeta_{ \beta \alpha}\hat{\Lambda}^{(\eta)}_{\beta \alpha}\right)}{\sqrt{2}|\zeta_{\alpha \beta}|}, \label{eq:G_11} \\
    &\hat{\boldsymbol{G}}^{(\eta)}_{\mu_{-}}:=\mi\frac{\left(\me^{-\mi \vec{k}_{0}\cdot \vec{r}_{\eta}}\zeta_{\alpha \beta}\hat{\Lambda}^{(\eta)}_{\alpha \beta}-\me^{\mi \vec{k}_{0}\cdot \vec{r}_{\eta}}\zeta_{ \beta \alpha}\hat{\Lambda}^{(\eta)}_{\beta \alpha}\right)}{\sqrt{2}|\zeta_{\alpha \beta}|},
    \label{eq:G_22}.
\end{align}
From Eqs.\eqref{eq:G_11} and \eqref{eq:G_22} are properly defined provided that $|\zeta_{\alpha \beta}|\neq 0,\ \forall\ (\alpha,\beta)$. Now, a complete the set of $d^{2}$ LOOs, is obtained by including $N$ population operators $\hat{\boldsymbol{G}}^{(\eta)}_{\mu_{z}}:=\hat{\Lambda}^{(\eta)}_{\beta \beta}$, each being associated to a specific internal atomic level $\beta$.
\section{Phase dependent local operators}
In this section, we will show that the operators $\{\hat{\boldsymbol{G}}^{(\eta)}_{\mu_{+}},\hat{\boldsymbol{G}}^{(\eta)}_{\mu_{-}},\hat{\boldsymbol{G}}^{(\eta)}_{\mu_{z}}\}$ satisfy the conditions to be a complete LOOs basis. For now on, we will consider the mapping $\mu:=\{\alpha\beta\}$.
\begin{equation}
     \begin{cases}
    \hat{\boldsymbol{G}}^{(\eta)}_{\mu_{+}}= |\zeta_{\alpha \beta}|^{-1}\frac{1}{\sqrt{2}}\left(\me^{-\mi \vec{k}_{0}\cdot \vec{r}_{\eta}}\zeta_{\alpha \beta}\hat{\Lambda}^{(\eta)}_{\alpha \beta}+\me^{\mi \vec{k}_{0}\cdot \vec{r}_{\eta}}\zeta_{ \beta \alpha}\hat{\Lambda}^{(\eta)}_{\beta \alpha}\right), & \text{if}\ 1\leq \mu_{+}\leq [d(d-1)]/2  \ \ \text{with}\ 1\leq \alpha < \beta \leq d, \\
    \hat{\boldsymbol{G}}^{(\eta)}_{\mu_{-}}=|\zeta_{\alpha \beta}|^{-1}\frac{i}{\sqrt{2}}\left(\me^{-\mi \vec{k}_{0}\cdot \vec{r}_{\eta}}\zeta_{\alpha \beta}\hat{\Lambda}^{(\eta)}_{\alpha \beta}-\me^{\mi \vec{k}_{0}\cdot \vec{r}_{\eta}}\zeta_{ \beta \alpha}\hat{\Lambda}^{(\eta)}_{\beta \alpha}\right), & \text{if}\  [d(d-1)]/2 < \mu_{-}\leq d(d-1)  \ \ \text{with}\ 1\leq \alpha < \beta \leq d,\\
    \hat{\boldsymbol{G}}^{(\eta)}_{\mu_{z}}=\hat{\Lambda}^{(\eta)}_{\beta \beta}, & \text{if}\  d(d-1) < \mu_{z} \leq d^{2}  \ \ \text{with}\ 1\leq  \beta \leq d.
  \end{cases}
  \label{eq:mylocaoperators_app}
\end{equation}

Eq.\eqref{eq:mylocaoperators_app} corresponds to a multilevel density matrix basis iff the following conditions are fulfilled: $\text{tr}\left[\hat{\boldsymbol{G}}^{(\eta)}_{\mu_{+}}\hat{\boldsymbol{G}}^{(\eta)}_{\mu'_{+}}\right]=\delta_{\mu_{+},\mu'_{+}}$, $\text{tr}\left[\hat{\boldsymbol{G}}^{(\eta)}_{\mu_{+}}\hat{\boldsymbol{G}}^{(\eta)}_{\mu_{-}}\right]=0$, $\text{tr}\left[\hat{\boldsymbol{G}}^{(\eta)}_{\mu_{+}}\hat{\boldsymbol{G}}^{(\eta)}_{\mu_{z}}\right]=0$, $\text{tr}\left[\hat{\boldsymbol{G}}^{(\eta)}_{\mu_{-}}\hat{\boldsymbol{G}}^{(\eta)}_{\mu'_{-}}\right]=\delta_{\mu_{-},\mu'_{-}}$, $\text{tr}\left[\hat{\boldsymbol{G}}^{(\eta)}_{\mu_{-}}\hat{\boldsymbol{G}}^{(\eta)}_{\mu_{z}}\right]=0$ and $\text{tr}\left[\hat{\boldsymbol{G}}^{(\eta)}_{\mu_{z}}\hat{\boldsymbol{G}}^{(\eta)}_{\mu_{z}}\right]=\delta_{\mu_{z},\mu'_{z}}$, so:

\begin{proof}
\begin{align}
    \nonumber &\text{tr}\left[\hat{\boldsymbol{G}}^{(\eta)}_{\mu_{+}}\hat{\boldsymbol{G}}^{(\eta)}_{\mu'_{+}}\right]=\text{tr}\left[\frac{|\zeta_{\alpha \beta}|^{-1}|\zeta_{\alpha' \beta'}|^{-1}}{2}\left(e^{-i\vec{k}_{0}\cdot \vec{r}_{\eta}}\zeta_{\alpha \beta}\hat{\Lambda}^{(\eta)}_{\alpha \beta}+e^{i\vec{k}_{0}\cdot \vec{r}_{\eta}}\zeta_{ \beta \alpha}\hat{\Lambda}^{(\eta)}_{\beta \alpha}\right)\left(e^{-i\vec{k}_{0}\cdot \vec{r}_{\eta}}\zeta_{\alpha' \beta'}\hat{\Lambda}^{(\eta)}_{\alpha' \beta'}+e^{i\vec{k}_{0}\cdot \vec{r}_{\eta}}\zeta_{ \beta' \alpha'}\hat{\Lambda}^{(\eta)}_{\beta' \alpha'}\right)\right]\\
    &\nonumber=\frac{|\zeta_{\alpha \beta}|^{-1}|\zeta_{\alpha' \beta'}|^{-1}}{2}e^{-i\vec{k}_{0}\cdot \vec{r}_{\eta}}e^{-i\vec{k}_{0}\cdot \vec{r}_{\eta}}\zeta_{\alpha \beta}\zeta_{\alpha' \beta'}\text{tr}\left[\hat{\Lambda}^{(\eta)}_{\alpha \beta}\hat{\Lambda}^{(\eta)}_{\alpha' \beta'}\right]+\frac{|\zeta_{\alpha \beta}|^{-1}|\zeta_{\alpha' \beta'}|^{-1}}{2}\zeta_{\alpha \beta}\zeta_{\beta'\alpha'}\text{tr}\left[\hat{\Lambda}^{(\eta)}_{\alpha \beta}\hat{\Lambda}^{(\eta)}_{\beta'\alpha'}\right]\\
    &\nonumber+\frac{|\zeta_{\alpha \beta}|^{-1}|\zeta_{\alpha' \beta'}|^{-1}}{2}e^{i\vec{k}_{0}\cdot \vec{r}_{\eta}}e^{i\vec{k}_{0}\cdot \vec{r}_{\eta}}\zeta_{\beta \alpha}\zeta_{\beta'\alpha'}\text{tr}\left[\hat{\Lambda}^{(\eta)}_{\beta\alpha }\hat{\Lambda}^{(\eta)}_{\beta'\alpha'}\right]+\frac{|\zeta_{\alpha \beta}|^{-1}|\zeta_{\alpha' \beta'}|^{-1}}{2}\zeta_{\beta \alpha}\zeta_{\alpha'\beta'}\text{tr}\left[\hat{\Lambda}^{(\eta)}_{\beta\alpha }\hat{\Lambda}^{(\eta)}_{\alpha'\beta'}\right]\\
    &\nonumber=\frac{|\zeta_{\alpha \beta}|^{-1}|\zeta_{\alpha' \beta'}|^{-1}}{2}\left(e^{-i\vec{k}_{0}\cdot \vec{r}_{\eta}}e^{-i\vec{k}_{0}\cdot \vec{r}_{\eta}}\zeta_{\alpha\beta}\zeta_{\alpha'\beta'}\delta_{\alpha \beta '}\delta_{\beta \alpha'}+\zeta_{\alpha\beta }\zeta_{\beta'\alpha'}\delta_{\alpha \alpha'}\delta_{\beta\beta'}\right)\\
    &\nonumber+\frac{|\zeta_{\alpha \beta}|^{-1}|\zeta_{\alpha' \beta'}|^{-1}}{2}\left(e^{i\vec{k}_{0}\cdot \vec{r}_{\eta}}e^{i\vec{k}_{0}\cdot \vec{r}_{\eta}}\zeta_{\beta\alpha }\zeta_{\beta'\alpha'}\delta_{\beta \alpha'}\delta_{\alpha \beta'}+\zeta_{\beta \alpha}\zeta_{\alpha'\beta'}\delta_{\beta\beta'}\delta_{\alpha \alpha'}\right)\\
    &=\delta_{\mu_{+}\mu'_{+}}.
\end{align}

\begin{align}
    \nonumber &\text{tr}\left[\hat{\boldsymbol{G}}^{(\eta)}_{\mu_{+}}\hat{\boldsymbol{G}}^{(\eta)}_{\mu_{-}}\right]=i\text{tr}\left[\frac{|\zeta_{\alpha \beta}|^{-1}|\zeta_{\alpha' \beta'}|^{-1}}{2}\left(e^{-i\vec{k}_{0}\cdot \vec{r}_{\eta}}\zeta_{\alpha \beta}\hat{\Lambda}^{(\eta)}_{\alpha \beta}+e^{i\vec{k}_{0}\cdot \vec{r}_{\eta}}\zeta_{ \beta \alpha}\hat{\Lambda}^{(\eta)}_{\beta \alpha}\right)\left(e^{-i\vec{k}_{0}\cdot \vec{r}_{\eta}}\zeta_{\alpha' \beta'}\hat{\Lambda}^{(\eta)}_{\alpha' \beta'}-e^{i\vec{k}_{0}\cdot \vec{r}_{\eta}}\zeta_{ \beta' \alpha'}\hat{\Lambda}^{(\eta)}_{\beta' \alpha'}\right)\right]\\
    &\nonumber=i\frac{|\zeta_{\alpha \beta}|^{-1}|\zeta_{\alpha' \beta'}|^{-1}}{2}e^{-i\vec{k}_{0}\cdot \vec{r}_{\eta}}e^{-i\vec{k}_{0}\cdot \vec{r}_{\eta}}\zeta_{\alpha \beta}\zeta_{\alpha' \beta'}\text{tr}\left[\hat{\Lambda}^{(\eta)}_{\alpha \beta}\hat{\Lambda}^{(\eta)}_{\alpha' \beta'}\right]-i\frac{|\zeta_{\alpha \beta}|^{-1}|\zeta_{\alpha' \beta'}|^{-1}}{2}\zeta_{\alpha \beta}\zeta_{\beta'\alpha'}\text{tr}\left[\hat{\Lambda}^{(\eta)}_{\alpha \beta}\hat{\Lambda}^{(\eta)}_{\beta'\alpha'}\right]\\
    &\nonumber-i\frac{|\zeta_{\alpha \beta}|^{-1}|\zeta_{\alpha' \beta'}|^{-1}}{2}e^{i\vec{k}_{0}\cdot \vec{r}_{\eta}}e^{i\vec{k}_{0}\cdot \vec{r}_{\eta}}\zeta_{\beta \alpha}\zeta_{\beta'\alpha'}\text{tr}\left[\hat{\Lambda}^{(\eta)}_{\beta\alpha }\hat{\Lambda}^{(\eta)}_{\beta'\alpha'}\right]+i\frac{|\zeta_{\alpha \beta}|^{-1}|\zeta_{\alpha' \beta'}|^{-1}}{2}\zeta_{\beta \alpha}\zeta_{\alpha'\beta'}\text{tr}\left[\hat{\Lambda}^{(\eta)}_{\beta\alpha }\hat{\Lambda}^{(\eta)}_{\alpha'\beta'}\right]\\
    &\nonumber=i\frac{|\zeta_{\alpha \beta}|^{-1}|\zeta_{\alpha' \beta'}|^{-1}}{2}\left(e^{-i\vec{k}_{0}\cdot \vec{r}_{\eta}}e^{-i\vec{k}_{0}\cdot \vec{r}_{\eta}}\zeta_{\alpha\beta}\zeta_{\alpha'\beta'}\delta_{\alpha \beta '}\delta_{\beta \alpha'}-\zeta_{\alpha\beta }\zeta_{\beta'\alpha'}\delta_{\alpha \alpha'}\delta_{\beta\beta'}\right)\\
    &\nonumber+i\frac{|\zeta_{\alpha \beta}|^{-1}|\zeta_{\alpha' \beta'}|^{-1}}{2}\left(-e^{i\vec{k}_{0}\cdot \vec{r}_{\eta}}e^{i\vec{k}_{0}\cdot \vec{r}_{\eta}}\zeta_{\beta\alpha }\zeta_{\beta'\alpha'}\delta_{\beta \alpha'}\delta_{\alpha \beta'}+\zeta_{\beta \alpha}\zeta_{\alpha'\beta'}\delta_{\beta\beta'}\delta_{\alpha \alpha'}\right)\\
    &=0.
\end{align}

\begin{align}
    \nonumber &\text{tr}\left[\hat{\boldsymbol{G}}^{(\eta)}_{\mu_{+}}\hat{\boldsymbol{G}}^{(\eta)}_{\mu_{z}}\right]=\text{tr}\left[\frac{|\zeta_{\alpha \beta}|^{-1}}{\sqrt{2}}\left(e^{-i\vec{k}_{0}\cdot \vec{r}_{\eta}}\zeta_{\alpha \beta}\hat{\Lambda}^{(\eta)}_{\alpha \beta}+e^{i\vec{k}_{0}\cdot \vec{r}_{\eta}}\zeta_{ \beta \alpha}\hat{\Lambda}^{(\eta)}_{\beta \alpha}\right)\left(\hat{\Lambda}^{(\eta)}_{\beta' \beta'}\right)\right]\\
    &\nonumber=\frac{|\zeta_{\alpha \beta}|^{-1}}{\sqrt{2}}e^{-i\vec{k}_{0}\cdot \vec{r}_{\eta}}\zeta_{\alpha \beta}\text{tr}\left[\hat{\Lambda}^{(\eta)}_{\alpha \beta}\hat{\Lambda}^{(\eta)}_{\beta' \beta'}\right]+\frac{|\zeta_{\alpha \beta}|^{-1}}{\sqrt{2}}e^{i\vec{k}_{0}\cdot \vec{r}_{\eta}}\zeta_{\beta \alpha}\text{tr}\left[\hat{\Lambda}^{(\eta)}_{\beta \alpha}\hat{\Lambda}^{(\eta)}_{\beta'\beta'}\right]\\
    &\nonumber=\frac{|\zeta_{\alpha \beta}|^{-1}}{\sqrt{2}}\left(e^{-i\vec{k}_{0}\cdot \vec{r}_{\eta}}\zeta_{\alpha\beta}\delta_{\alpha \beta '}\delta_{\beta \beta'}+e^{i\vec{k}_{0}\cdot \vec{r}_{\eta}}\zeta_{\beta\alpha }\delta_{\alpha \beta'}\delta_{\beta\beta'}\right)\\
    &=0.
\end{align}

\begin{align}
    \nonumber &\text{tr}\left[\hat{\boldsymbol{G}}^{(\eta)}_{\mu_{-}}\hat{\boldsymbol{G}}^{(\eta)}_{\mu'_{-}}\right]=-\text{tr}\left[\frac{|\zeta_{\alpha \beta}|^{-1}|\zeta_{\alpha' \beta'}|^{-1}}{2}\left(e^{-i\vec{k}_{0}\cdot \vec{r}_{\eta}}\zeta_{\alpha \beta}\hat{\Lambda}^{(\eta)}_{\alpha \beta}-e^{i\vec{k}_{0}\cdot \vec{r}_{\eta}}\zeta_{ \beta \alpha}\hat{\Lambda}^{(\eta)}_{\beta \alpha}\right)\left(e^{-i\vec{k}_{0}\cdot \vec{r}_{\eta}}\zeta_{\alpha' \beta'}\hat{\Lambda}^{(\eta)}_{\alpha' \beta'}-e^{i\vec{k}_{0}\cdot \vec{r}_{\eta}}\zeta_{ \beta' \alpha'}\hat{\Lambda}^{(\eta)}_{\beta' \alpha'}\right)\right]\\
    &\nonumber=-\frac{|\zeta_{\alpha \beta}|^{-1}|\zeta_{\alpha' \beta'}|^{-1}}{2}e^{-i\vec{k}_{0}\cdot \vec{r}_{\eta}}e^{-i\vec{k}_{0}\cdot \vec{r}_{\eta}}\zeta_{\alpha \beta}\zeta_{\alpha' \beta'}\text{tr}\left[\hat{\Lambda}^{(\eta)}_{\alpha \beta}\hat{\Lambda}^{(\eta)}_{\alpha' \beta'}\right]+\frac{|\zeta_{\alpha \beta}|^{-1}|\zeta_{\alpha' \beta'}|^{-1}}{2}\zeta_{\alpha \beta}\zeta_{\beta'\alpha'}\text{tr}\left[\hat{\Lambda}^{(\eta)}_{\alpha \beta}\hat{\Lambda}^{(\eta)}_{\beta'\alpha'}\right]\\
    &\nonumber-\frac{|\zeta_{\alpha \beta}|^{-1}|\zeta_{\alpha' \beta'}|^{-1}}{2}e^{i\vec{k}_{0}\cdot \vec{r}_{\eta}}e^{i\vec{k}_{0}\cdot \vec{r}_{\eta}}\zeta_{\beta \alpha}\zeta_{\beta'\alpha'}\text{tr}\left[\hat{\Lambda}^{(\eta)}_{\beta\alpha }\hat{\Lambda}^{(\eta)}_{\beta'\alpha'}\right]+\frac{|\zeta_{\alpha \beta}|^{-1}|\zeta_{\alpha' \beta'}|^{-1}}{2}\zeta_{\beta \alpha}\zeta_{\alpha'\beta'}\text{tr}\left[\hat{\Lambda}^{(\eta)}_{\beta\alpha }\hat{\Lambda}^{(\eta)}_{\alpha'\beta'}\right]\\
    &\nonumber=\frac{|\zeta_{\alpha \beta}|^{-1}|\zeta_{\alpha' \beta'}|^{-1}}{2}\left(-e^{-i\vec{k}_{0}\cdot \vec{r}_{\eta}}e^{-i\vec{k}_{0}\cdot \vec{r}_{\eta}}\zeta_{\alpha\beta}\zeta_{\alpha'\beta'}\delta_{\alpha \beta '}\delta_{\beta \alpha'}+\zeta_{\alpha\beta }\zeta_{\beta'\alpha'}\delta_{\alpha \alpha'}\delta_{\beta\beta'}\right)\\
    &\nonumber+\frac{|\zeta_{\alpha \beta}|^{-1}|\zeta_{\alpha' \beta'}|^{-1}}{2}\left(-e^{i\vec{k}_{0}\cdot \vec{r}_{\eta}}e^{i\vec{k}_{0}\cdot \vec{r}_{\eta}}\zeta_{\beta\alpha }\zeta_{\beta'\alpha'}\delta_{\beta \alpha'}\delta_{\alpha \beta'}+\zeta_{\beta \alpha}\zeta_{\alpha'\beta'}\delta_{\beta\beta'}\delta_{\alpha \alpha'}\right)\\
    &=\delta_{\mu_{-}\mu'_{-}}.
\end{align}

\begin{align}
    \nonumber &\text{tr}\left[\hat{\boldsymbol{G}}^{(\eta)}_{\mu_{-}}\hat{\boldsymbol{G}}^{(\eta)}_{\mu_{z}}\right]=i\text{tr}\left[\frac{|\zeta_{\alpha \beta}|^{-1}}{\sqrt{2}}\left(e^{-i\vec{k}_{0}\cdot \vec{r}_{\eta}}\zeta_{\alpha \beta}\hat{\Lambda}^{(\eta)}_{\alpha \beta}-e^{i\vec{k}_{0}\cdot \vec{r}_{\eta}}\zeta_{ \beta \alpha}\hat{\Lambda}^{(\eta)}_{\beta \alpha}\right)\left(\hat{\Lambda}^{(\eta)}_{\beta' \beta'}\right)\right]\\
    &\nonumber=i\frac{|\zeta_{\alpha \beta}|^{-1}}{\sqrt{2}}e^{-i\vec{k}_{0}\cdot \vec{r}_{\eta}}\zeta_{\alpha \beta}\text{tr}\left[\hat{\Lambda}^{(\eta)}_{\alpha \beta}\hat{\Lambda}^{(\eta)}_{\beta' \beta'}\right]-i\frac{|\zeta_{\alpha \beta}|^{-1}}{\sqrt{2}}e^{i\vec{k}_{0}\cdot \vec{r}_{\eta}}\zeta_{\beta \alpha}\text{tr}\left[\hat{\Lambda}^{(\eta)}_{\beta \alpha}\hat{\Lambda}^{(\eta)}_{\beta'\beta'}\right]\\
    &\nonumber=i\frac{|\zeta_{\alpha \beta}|^{-1}}{\sqrt{2}}\left(e^{-i\vec{k}_{0}\cdot \vec{r}_{\eta}}\zeta_{\alpha\beta}\delta_{\alpha \beta '}\delta_{\beta \beta'}-e^{i\vec{k}_{0}\cdot \vec{r}_{\eta}}\zeta_{\beta\alpha }\delta_{\alpha \beta'}\delta_{\beta\beta'}\right)\\
    &=0.
\end{align}
\begin{align}
    \text{tr}\left[\hat{\boldsymbol{G}}^{(\eta)}_{\mu_{z}}\hat{\boldsymbol{G}}^{(\eta)}_{\mu'_{z}}\right]=\text{tr}\left[\hat{\Lambda}^{(\eta)}_{\beta \beta}\hat{\Lambda}^{(\eta)}_{\beta' \beta'}\right]=\delta_{\beta \beta'}\delta_{\beta \beta'}=\delta_{\mu_{z},\mu'_{z}}.
\end{align}
\end{proof}
\section{Result 1: First inequality}

\begin{proof}
In order to proof the result concerning the first inequality in the main text, let us consider three arbitrary collective operators.
 \begin{align}
     \hat{A}_{\mu_{+}}=\sum_{\eta=1}^{N} \hat{\boldsymbol{G}}^{(\eta)}_{\mu_{+}},\ \hat{B}_{\mu_{-}}=\sum_{\eta=1}^{N} \hat{\boldsymbol{G}}^{(\eta)}_{\mu_{-}}, \ \hat{C}_{\mu_{z}}=\sum_{\eta=1}^{N} \hat{\boldsymbol{G}}^{(\eta)}_{\mu_{z}}.
 \end{align}
 
 Using the variance concavity relation for a separable state $\hat{\varrho}_{\text{sep}}=\bigotimes_{\eta=1}^{N}\hat{\varrho}_{\eta}$, we get;

\begin{align}
    &(\Delta \hat{A}_{\mu_{+}})^{2}\geq \sum_{\eta=1}^{N}\left(\Delta \hat{\boldsymbol{G}}^{(\eta)}_{\mu_{+}}\right)^{2}_{\eta},\\
    &(\Delta \hat{B}_{\mu_{-}})^{2}\geq  \sum_{\eta=1}^{N}\left(\Delta \hat{\boldsymbol{G}}^{(\eta)}_{\mu_{-}}\right)^{2}_{\eta},\\
    &(\Delta \hat{C}_{\mu_{z}})^{2}\geq  \sum_{\eta=1}^{N}\left(\Delta \hat{\boldsymbol{G}}^{(\eta)}_{\mu_{z}}\right)^{2}_{\eta}.
\end{align}
so we obtain the relation
\begin{align}
   \nonumber \sum_{\mu_{+}}(\Delta \hat{A}_{\mu_{+}})^{2}+\sum_{\mu_{-}}(\Delta \hat{B}_{\mu_{-}})^{2}+\sum_{\mu_{z}}(\Delta \hat{C}_{\mu_{z}})^{2}&\geq \sum_{\mu_{+}}\sum_{\eta=1}^{N}\left(\Delta \hat{\boldsymbol{G}}^{(\eta)}_{\mu_{+}}\right)^{2}_{\eta}+\sum_{\mu_{-}}\sum_{\eta=1}^{N}\left(\Delta \hat{\boldsymbol{G}}^{(\eta)}_{\mu_{-}}\right)^{2}_{\eta}+\sum_{\mu_{z}}\sum_{\eta=1}^{N}\left(\Delta \hat{\boldsymbol{G}}^{(\eta)}_{\mu_{z}}\right)^{2}_{\eta}\\
    &\nonumber =\sum_{\eta=1}^{N}\left\langle \sum_{\mu_{+}}\left(\hat{\boldsymbol{G}}^{(\eta)}_{\mu_{+}}\right)^{2} +\sum_{\mu_{-}}\left(\hat{\boldsymbol{G}}^{(\eta)}_{\mu_{-}}\right)^{2} + \sum_{\mu_{z}}\left(\hat{\boldsymbol{G}}^{(\eta)}_{\mu_{z}}\right)^{2}\right\rangle\\
    &-\sum_{\eta=1}^{N}\left(\sum_{\mu_{+}}\left\langle \hat{\boldsymbol{G}}^{(\eta)}_{\mu_{+}}\right\rangle^{2}+\sum_{\mu_{-}}\left\langle \hat{\boldsymbol{G}}^{(\eta)}_{\mu_{-}}\right\rangle^{2}+\sum_{\mu_{z}}\left\langle \hat{\boldsymbol{G}}^{(\eta)}_{\mu_{z}}\right\rangle^{2}\right).
\end{align}
Applying the properties for a general set of LOOs $\sum_{m=1}^{d^{2}}\hat{\boldsymbol{G}}_{m}\hat{\boldsymbol{G}}_{m}=\id_d$ and $\sum_{m=1}^{d^{2}}\langle \hat{\boldsymbol{G}}_{m}\rangle^{2}\leq 1$ (purity), the final result reads as follows
\begin{align}
    \sum_{\mu_{+}}(\Delta \hat{A}_{\mu_{+}})^{2}+\sum_{\mu_{-}}(\Delta \hat{B}_{\mu_{-}})^{2}+\sum_{\mu_{z}}(\Delta \hat{C}_{\mu_{z}})^{2}\geq (d-1)N.
\end{align}
\end{proof}
The previous inequality also holds for a general separable state of the form $\hat{\varrho}_{\text{sep}}=\sum_{l=1}^{L}p_{l}\bigotimes_{\eta=1}^{N}\hat{\varrho}^{(l)}_{\eta}$, by using the variance's concavity relation.
\section{Result 1: Second inequality}
\begin{proof}

 Considering again three arbitrary operators.
\begin{align}
    &\hat{A}_{\mu_{+}}=\sum_{\eta=1}^{N}\hat{\boldsymbol{G}}^{(\eta)}_{\mu_{+}}, \ \hat{B}_{\mu_{-}}=\sum_{\eta=1}^{N}\hat{\boldsymbol{G}}^{(\eta)}_{\mu_{-}},\ \ \hat{C}_{\mu_{z}}=\sum_{\eta=1}^{N}\hat{\boldsymbol{G}}^{(\eta)}_{\mu_{z}}.
\end{align}
Now we define the modified second moment operator.
\begin{align}
    \langle \bar{A}^{2}_{\mu_{k}}\rangle=\langle \hat{A}^{2}_{\mu_{k}}\rangle-\left\langle  \sum_{\eta=1}^{N}\left(\hat{\boldsymbol{G}}^{(\eta)}_{\mu_{k}}\right)^{2}\right\rangle.
\end{align}
and modified variance
\begin{align}
    (\Delta \bar{A}_{\mu_{k}})^{2}=(\Delta \hat{A}_{\mu_{k}})^{2}-\left\langle  \sum_{\eta=1}^{N}\left(\hat{\boldsymbol{G}}^{(\eta)}_{\mu_{k}}\right)^{2}\right\rangle.
\end{align}
For a separable state $\hat{\varrho}_{\text{sep}}=\bigotimes_{\kappa=1}^{N}\hat{\varrho}^{(\kappa)}$, we obtain the following.
\begin{align}
    \nonumber \langle \bar{A}^{2}_{\mu_{k}}\rangle &=\langle \hat{A}^{2}_{\mu_{k}}\rangle-\left\langle  \sum_{\eta=1}^{N}\left(\hat{\boldsymbol{G}}^{(\eta)}_{\mu_{k}}\right)^{2}\right\rangle\\
    &\nonumber=\left \langle \sum_{\eta=1}^{N}\sum_{\eta'\neq \eta}^{N}\hat{\boldsymbol{G}}^{(\eta)}_{\mu_{k}}\hat{\boldsymbol{G}}^{(\eta')}_{\mu_{k}} \right\rangle+\sum_{\eta=1}^{N}\left\langle\hat{\boldsymbol{G}}^{(\eta)}_{\mu_{k}}\right\rangle^{2}-\sum_{\eta=1}^{N}\left\langle\hat{\boldsymbol{G}}^{(\eta)}_{\mu_{k}}\right\rangle^{2}\\
    &=\langle \hat{A}_{\mu_{k}}\rangle^{2}-\sum_{\eta=1}^{N}\left\langle\hat{\boldsymbol{G}}^{(\eta)}_{\mu_{k}}\right\rangle^{2}.
\end{align}
Similarly, the modified variance can be simplified as:
\begin{align}
    \nonumber (\Delta \bar{A}_{\mu_{k}})^{2}&=(\Delta \hat{A}_{\mu_{k}})^{2}-\left\langle  \sum_{\eta=1}^{N}\left(\hat{\boldsymbol{G}}^{(\eta)}_{\mu_{k}}\right)^{2}\right\rangle\\
    &\nonumber=\langle \hat{A}^{2}_{\mu_{k}}\rangle-\langle\hat{A}_{\mu_{k}}\rangle^{2}-\left\langle  \sum_{\eta=1}^{N}\left(\hat{\boldsymbol{G}}^{(\eta)}_{\mu_{k}}\right)^{2}\right\rangle\\
    &\nonumber =\langle \hat{A}_{\mu_{k}}\rangle^{2}-\sum_{\eta=1}^{N}\left\langle\hat{\boldsymbol{G}}^{(\eta)}_{\mu_{k}}\right\rangle^{2}-\langle\hat{A}_{\mu_{k}}\rangle^{2}\\
    &=-\sum_{\eta=1}^{N}\left\langle\hat{\boldsymbol{G}}^{(\eta)}_{\mu_{k}}\right\rangle^{2}.
\end{align}
Using the previous expression we get:
\begin{align}
    \nonumber &-\sum_{\mu_{+}}\langle \bar{{A}}^{2}_{\mu_{+}}\rangle- \sum_{\mu_{-}}\langle \bar{{B}}^{2}_{\mu_{-}}\rangle+(N-1)\sum_{\mu_{z}}(\Delta \bar{{C}}_{\mu_{z}})^{2}=-\sum_{\mu_{+}}\left(\langle \hat{A}_{\mu_{+}}\rangle^{2}-\sum_{\eta=1}^{N}\left\langle\hat{\boldsymbol{G}}^{(\eta)}_{\mu_{+}}\right\rangle^{2}\right)\\
    &\nonumber-\sum_{\mu_{-}}\left(\langle \hat{B}_{\mu_{-}}\rangle^{2}-\sum_{\eta=1}^{N}\left\langle\hat{\boldsymbol{G}}^{(\eta)}_{\mu_{-}}\right\rangle^{2}\right)-(N-1)\sum_{\mu_{z}}\sum_{\eta=1}^{N}\left\langle\hat{\boldsymbol{G}}^{(\eta)}_{\mu_{z}}\right\rangle^{2}\\
&\nonumber=\sum_{\eta=1}^{N}\left(\sum_{\mu_{+}}\left\langle\hat{\boldsymbol{G}}^{(\eta)}_{\mu_{+}}\right\rangle^{2}+\sum_{\mu_{-}}\left\langle\hat{\boldsymbol{G}}^{(\eta)}_{\mu_{-}}\right\rangle^{2}+\sum_{\mu_{z}}\left\langle\hat{\boldsymbol{G}}^{(\eta)}_{\mu_{z}}\right\rangle^{2}\right)\\
    &\nonumber -\sum_{\mu_{+}}\langle \hat{A}_{\mu_{+}}\rangle^{2}-\sum_{\mu_{-}}\langle \hat{B}_{\mu_{-}}\rangle^{2}-N\sum_{\mu_{z}}\sum_{\eta=1}^{N}\left\langle\hat{\boldsymbol{G}}^{(\eta)}_{\mu_{z}}\right\rangle^{2}\\
    &\nonumber \geq \sum_{\eta=1}^{N}\left(\sum_{\mu_{+}}\left\langle\hat{\boldsymbol{G}}^{(\eta)}_{\mu_{+}}\right\rangle^{2}+\sum_{\mu_{-}}\left\langle\hat{\boldsymbol{G}}^{(\eta)}_{\mu_{-}}\right\rangle^{2}+\sum_{\mu_{z}}\left\langle\hat{\boldsymbol{G}}^{(\eta)}_{\mu_{z}}\right\rangle^{2}\right)\\
    &-N\sum_{\mu_{+}}\sum_{\eta}^{N}\langle \hat{\boldsymbol{G}}^{(\eta)}_{\mu_{+}} \rangle^{2}-N\sum_{\mu_{-}}\sum_{\eta}^{N}\langle \hat{\boldsymbol{G}}^{(\eta)}_{\mu_{-}} \rangle^{2}-N\sum_{\mu_{z}}\sum_{\eta=1}^{N}\langle\hat{\boldsymbol{G}}^{(\eta)}_{\mu_{z}}\rangle^{2}\\
    &=-N(N-1).
\end{align}
where we have used the properties (which follow from Cauchy–Schwarz inequality)
\begin{align}
    &\langle \hat{A}_{\mu_{+}} \rangle^{2}\leq N\sum_{\eta}^{N}\langle \hat{\boldsymbol{G}}^{(\eta)}_{\mu_{+}} \rangle^{2},\\
    &\langle \hat{B}_{\mu_{-}} \rangle^{2}\leq N\sum_{\eta}^{N}\langle \hat{\boldsymbol{G}}^{(\eta)}_{\mu_{-}} \rangle^{2}.
\end{align}
Finally, we arrive to the expression representing the second inequality.
\begin{align}
   (N-1)\sum_{\mu_{z}}(\Delta \bar{{C}}_{\mu_{z}})^{2} -\sum_{\mu_{+}}\langle \bar{{A}}^{2}_{\mu_{+}}\rangle- \sum_{\mu_{-}}\langle \bar{{B}}^{2}_{\mu_{-}}\rangle\geq -N(N-1).
\end{align}
\end{proof}

Note that the above inequality applies for any permutation of the operators $\{\bar{{A}}_{\mu_{+}},\bar{{B}}_{\mu_{-}},\bar{{C}}_{\mu_{z}}\}$, i,e:
\begin{align}
   &(N-1)\sum_{\mu_{+}}(\Delta \bar{{A}}_{\mu_{+}})^{2} -\sum_{\mu_{z}}\langle \bar{{C}}^{2}_{\mu_{z}}\rangle- \sum_{\mu_{-}}\langle \bar{{B}}^{2}_{\mu_{-}}\rangle\geq -N(N-1),\\
   &(N-1)\sum_{\mu_{-}}(\Delta \bar{{B}}_{\mu_{-}})^{2} -\sum_{\mu_{z}}\langle \bar{{C}}^{2}_{\mu_{z}}\rangle- \sum_{\mu_{+}}\langle \bar{{A}}^{2}_{\mu_{+}}\rangle\geq -N(N-1).
   \label{eq:app_second}
\end{align}

\section{Result 1: Third inequality}
\begin{proof}
Following a similar calculation to the second inequality, we get:
\begin{align}
   &(N-1)\left(\sum_{\mu_{z}}(\Delta \bar{{C}}_{\mu_{z}})^{2}+\sum_{\mu_{-}}(\Delta \bar{{B}}_{\mu_{-}})^{2}\right) -\sum_{\mu_{+}}\langle \bar{{A}}^{2}_{\mu_{+}}\rangle\geq -N(N-1),\\
   &(N-1)\left(\sum_{\mu_{z}}(\Delta \bar{{C}}_{\mu_{z}})^{2}+\sum_{\mu_{+}}(\Delta \bar{{A}}_{\mu_{+}})^{2}\right) -\sum_{\mu_{-}}\langle \bar{{B}}^{2}_{\mu_{-}}\rangle\geq -N(N-1),\\
   &(N-1)\left(\sum_{\mu_{+}}(\Delta \bar{{A}}_{\mu_{+}})^{2}+\sum_{\mu_{-}}(\Delta \bar{{B}}_{\mu_{-}})^{2}\right) -\sum_{\mu_{z}}\langle \bar{{C}}^{2}_{\mu_{z}}\rangle\geq -N(N-1).
\end{align}
\end{proof}

\section{ Remark 2: On light polarization independence}
\begin{proof}

Having in mind that the population operator $\hat{\boldsymbol{G}}^{(\eta)}_{\mu_{z}}$ is independent of $\zeta_{\alpha\beta}$, we notice that $w_{1,\boldsymbol{k}_{0}}$ is directly related to the quantity:

    \begin{align}
    \nonumber&(\Delta \hat{\boldsymbol{X}}_{\alpha\beta})^{2}+(\Delta \hat{\boldsymbol{Y}}_{\alpha\beta})^{2}=\sum_{\eta=1}^{N}\left(\langle\hat{\Lambda}^{(\eta)}_{\alpha\alpha}\rangle+\langle\hat{\Lambda}^{(\eta)}_{\beta\beta}\rangle\right)+\sum_{\eta=1}^{N}\sum_{\eta'\neq \eta}^{N}
    \left(\me^{\mi\vec{k}_{0}\cdot(\vec{r}_{\eta}-\vec{r}_{\eta'})}\langle\hat{\Lambda}^{(\eta)}_{\beta\alpha}\hat{\Lambda}^{(\eta')}_{\alpha\beta}\rangle+\me^{-\mi\vec{k}_{0}\cdot(\vec{r}_{\eta}-\vec{r}_{\eta'})}\langle\hat{\Lambda}^{(\eta)}_{\alpha\beta}\hat{\Lambda}^{(\eta')}_{\beta\alpha}\rangle\right)\\
    &-\sum_{\eta=1}^{N}\left(\langle\hat{\Lambda}^{(\eta)}_{\alpha\beta}\rangle\langle\hat{\Lambda}^{(\eta)}_{\beta\alpha}\rangle+\langle\hat{\Lambda}^{(\eta)}_{\beta\alpha}\rangle\langle\hat{\Lambda}^{(\eta)}_{\alpha\beta}\rangle\right)-\sum_{\eta=1}^{N}\sum_{\eta'\neq \eta}^{N}\left(\me^{-\mi\vec{k}_{0}\cdot(\vec{r}_{\eta}-\vec{r}_{\eta'})}\langle\hat{\Lambda}^{(\eta)}_{\alpha\beta}\rangle\langle\hat{\Lambda}^{(\eta')}_{\beta\alpha}\rangle+\me^{\mi\vec{k}_{0}\cdot(\vec{r}_{\eta}-\vec{r}_{\eta'})}\langle\hat{\Lambda}^{(\eta)}_{\beta\alpha}\rangle\langle\hat{\Lambda}^{(\eta')}_{\alpha\beta}\rangle\right).
\end{align}
The proof of Remark \ref{remark:2} ends with the observation that the last expression is independent of $\zeta_{\alpha\beta}$.

Using the same approach as before, it is possible to show that the following  two quantities $\{w^{\bar{\boldsymbol{Z}},\bar{\boldsymbol{X}},\bar{\boldsymbol{Y}}}_{2,\vec{k}_{0}},w^{\bar{\boldsymbol{X}},\bar{\boldsymbol{Y}},\bar{\boldsymbol{Z}}}_{3,\vec{k}_{0}}\}$, are also polarization independent. The last statement follows by noticing that both quantities can associated to the terms: $-\langle \bar{\boldsymbol{X}}^{2}_{\alpha\beta}\rangle- \langle \bar{\boldsymbol{Y}}^{2}_{\alpha\beta}\rangle$ and $(\Delta \bar{\boldsymbol{X}}_{\alpha\beta})^{2}+(\Delta \bar{\boldsymbol{Y}}_{\alpha\beta})^{2}$ respectively.

\end{proof}


\section{Symmetric state with white noise}
Let us consider the general operators
\begin{align}
    &\hat{\boldsymbol{X}}_{\mu_{+}}=\sum_{\eta=1}^{N}\hat{\boldsymbol{G}}^{(\eta)}_{\mu_{+}},\ \hat{\boldsymbol{Y}}_{\mu_{-}}=\sum_{\eta=1}^{N}\hat{\boldsymbol{G}}^{(\eta)}_{\mu_{-}},\ \hat{\boldsymbol{Z}}_{\mu_{z}}=\sum_{\eta=1}^{N}\hat{\boldsymbol{G}}^{(\eta)}_{\mu_{z}}.
\end{align}
and the quantum state:
\begin{align}
    \hat{\varrho}= p\frac{\id}{d^{N}}+(1-p)\ketbra{S_{N}}{S_{N}}.
\end{align}
with $\ket{S_{N}}$ the symmetric state defined in the main text. Then (with $d=N$) after some algebra we arrive to the expressions:
\begin{align}
    &\langle \hat{\boldsymbol{X}}_{\mu_{+}} \rangle=0,\\
    &\langle \hat{\boldsymbol{Y}}_{\mu_{-}} \rangle=0,\\
    &\langle \hat{\boldsymbol{Z}}_{\mu_{z}} \rangle =1.
\end{align}

For the higher-order momenta, we get

\begin{align}
 &\langle \hat{\boldsymbol{X}}^{2}_{\mu_{+}} \rangle =1+\frac{(1-p)}{N(N-1)}\left(\left|\sum_{\eta=1}^{N}\me^{\mi \vec{k}_{0}\cdot \vec{r}_{\eta}}\right|^{2}-N\right),\\
 &\langle \hat{\boldsymbol{Y}}^{2}_{\mu_{-}} \rangle=1+\frac{(1-p)}{N(N-1)}\left(\left|\sum_{\eta=1}^{N}\me^{\mi \vec{k}_{0}\cdot \vec{r}_{\eta}}\right|^{2}-N\right),\\
 &\langle \hat{\boldsymbol{Z}}^{2}_{\mu_{z}} \rangle=1+p\frac{(N-1)}{N}.
\end{align}

so $w_{1,\vec{k}_{0}}$ takes the form, (introducing the structure factor $S_{\vec{k}_{0}}=\left|\sum_{\eta=1}^{N}\me^{\mi \vec{k}_{0}\cdot \vec{r}_{\eta}}\right|^{2}$)
\begin{align}
    2\sum_{\mu_{1}=1}^{N(N-1)/2}\left[1+\frac{(1-p)}{N(N-1)}\left(S_{\vec{k}_{0}}-N\right)\right]+\sum_{\mu_{3}=1}^{N}p\frac{(N-1)}{N}\geq N(N-1).
\end{align}
From the last expression, after some simplifications, we obtain the condition (Eq.\eqref{eq:sym_p}) presented in the main text.

\section{Asymmetric state with white noise}
Let the asymmetric quantum state with white noise be
\begin{align}
    \hat{\varrho}= p\frac{\id}{d^{N}}+(1-p)\ketbra{A_{N}}{A_{N}}.
\end{align} 
Then, considering the case $d=N$, we get after some algebra:
\begin{align}
    &\langle \hat{\boldsymbol{X}}_{\mu_{+}} \rangle=0,\\
    &\langle \hat{\boldsymbol{Y}}_{\mu_{-}} \rangle=0,\\
    &\langle \hat{\boldsymbol{Z}}_{\mu_{z}} \rangle =1.
\end{align}

\begin{align}
   &\langle \hat{\boldsymbol{X}}^{2}_{\mu_{+}} \rangle =1-\frac{(1-p)}{N(N-1)}\left(\left|\sum_{\eta=1}^{N}\me^{\mi \vec{k}_{0}\cdot \vec{r}_{\eta}}\right|^{2}-N\right),\\
   &\langle \hat{\boldsymbol{Y}}^{2}_{\mu_{-}} \rangle =1-\frac{(1-p)}{N(N-1)}\left(\left|\sum_{\eta=1}^{N}\me^{\mi \vec{k}_{0}\cdot \vec{r}_{\eta}}\right|^{2}-N\right),\\
   &\langle \hat{\boldsymbol{Z}}^{2}_{\mu_{z}} \rangle=1+p\frac{(N-1)}{N}.
\end{align}

Now, the quantity $w_{1,\vec{k}_{0}}$  reads:
\begin{align}
    2\sum_{\mu_{1}=1}^{N(N-1)/2}\left[1-\frac{(1-p)}{N(N-1)}\left(S_{\vec{k}_{0}}-N\right)\right]+\sum_{\mu_{3}=1}^{N}p\frac{(N-1)}{N}\geq N(N-1).
\end{align}
From the last expression, the condition (Eq.\eqref{eq:asym_p}) for entanglement presented in the main text is obtained.

\section{W-like states}
The generalized W state for qudits is given by
\begin{align}
    \nonumber\ket{\text{W}_{d}}&=\frac{1}{\sqrt{(d-1)N}}\sum_{i=1}^{N}\sum_{j=2}^{d}\ket{1}^{(i-1)}\otimes\ket{j}_{i}\otimes\ket{1}^{(N-i)}\\
    &=\frac{1}{\sqrt{(d-1)N}}\sum_{j=2}^{d}(\ket{j1\hdots1}+\ket{1j\hdots1}\hdots+\ket{1\hdots1j}),
\end{align}
The expected value of the linear operators will take the form:
\begin{align}
    &\langle \hat{\boldsymbol{Z}}_{11} \rangle=\bra{\text{W}_{d}}\hat{\boldsymbol{Z}}_{11} \ket{\text{W}_{d}}=(N-1).
\end{align}
for $2\leq\beta\leq d$, we get
\begin{align}
    &\langle \hat{\boldsymbol{Z}}_{\beta\beta} \rangle=\bra{\text{W}_{d}}\hat{\boldsymbol{Z}}_{\beta\beta} \ket{\text{W}_{d}}=\frac{1}{d-1},\\
    &\langle \hat{\boldsymbol{X}}_{1\beta} \rangle=0,\\
    &\langle \hat{\boldsymbol{Y}}_{1\beta} \rangle=0.
\end{align}
For $\alpha\neq 1 < \beta$ and $\beta\geq 3$, we get
\begin{align}
    &\langle \hat{\boldsymbol{X}}_{\alpha \beta} \rangle=\frac{1}{\sqrt{2}|\zeta_{\alpha\beta}|(d-1)N}\sum_{\eta=1}^{N}\left(e^{-\mi \vec{k}_{0}\cdot \vec{r}_{\eta}}\zeta_{\alpha\beta}+e^{\mi \vec{k}_{0}\cdot \vec{r}_{\eta}}\zeta_{\beta\alpha}\right),\\
    &\langle \hat{\boldsymbol{Y}}_{\alpha \beta} \rangle=\frac{\mi}{\sqrt{2}|\zeta_{\alpha\beta}|(d-1)N}\sum_{\eta=1}^{N}\left(e^{-\mi \vec{k}_{0}\cdot \vec{r}_{\eta}}\zeta_{\alpha\beta}-e^{\mi \vec{k}_{0}\cdot \vec{r}_{\eta}}\zeta_{\beta\alpha}\right).
\end{align}
Similarly for the population operator
\begin{align}
    &\langle \hat{\boldsymbol{Z}}^{2}_{11} \rangle=\bra{\text{W}_{d}}\hat{\boldsymbol{Z}}^{2}_{11} \ket{\text{W}_{d}}=(N-1)+(N-1)(N-2)=(N-1)^{2}.
\end{align}
For the W state, it is possible to show that $\hat{\boldsymbol{X}}^{2}_{\alpha\beta}$ and $\hat{\boldsymbol{Y}}^{2}_{\alpha\beta}$ can be written as:
\begin{align}
     &\hat{\boldsymbol{X}}^{2}_{\alpha\beta}= \frac{1}{2}\sum_{\eta=1}^{N}(\Lambda^{(\eta)}_{\alpha\alpha}+\Lambda^{(\eta)}_{\beta\beta})+\frac{1}{2}\sum_{\eta=1}^{N}\sum_{\eta'\neq \eta}^{N}\left(\me^{-\mi \vec{k}_{0}\cdot \vec{r}_{\eta}}\me^{\mi \vec{k}_{0}\cdot \vec{r}_{\eta'}}\hat{\Lambda}^{(\eta)}_{\alpha\beta}\hat{\Lambda}^{(\eta')}_{\beta\alpha}+\me^{\mi \vec{k}_{0}\cdot \vec{r}_{\eta}}\me^{-\mi \vec{k}_{0}\cdot \vec{r}_{\eta'}}\hat{\Lambda}^{(\eta)}_{\beta\alpha}\hat{\Lambda}^{(\eta')}_{\alpha\beta}\right),\\
     &\hat{\boldsymbol{Y}}^{2}_{\alpha\beta}= \frac{1}{2}\sum_{\eta=1}^{N}(\Lambda^{(\eta)}_{\alpha\alpha}+\Lambda^{(\eta)}_{\beta\beta})+\frac{1}{2}\sum_{\eta=1}^{N}\sum_{\eta'\neq \eta}^{N}\left(\me^{-\mi \vec{k}_{0}\cdot \vec{r}_{\eta}}\me^{\mi \vec{k}_{0}\cdot \vec{r}_{\eta'}}\hat{\Lambda}^{(\eta)}_{\alpha\beta}\hat{\Lambda}^{(\eta')}_{\beta\alpha}+\me^{\mi \vec{k}_{0}\cdot \vec{r}_{\eta}}\me^{-\mi \vec{k}_{0}\cdot \vec{r}_{\eta'}}\hat{\Lambda}^{(\eta)}_{\beta\alpha}\hat{\Lambda}^{(\eta')}_{\alpha\beta}\right).
\end{align}
For $2\leq\beta\leq d$, we obtain the expressions:
\begin{align}
    &\langle \hat{\boldsymbol{Z}}^{2}_{\beta\beta} \rangle=\bra{\text{W}_{d}}\hat{\boldsymbol{Z}}^{2}_{\beta\beta} \ket{\text{W}_{d}}=\frac{1}{d-1},\\
    &\langle \hat{\boldsymbol{X}}^{2}_{1\beta} \rangle=\frac{(N-1)}{2}+\frac{1}{2(d-1)}+\frac{1}{2(d-1)N}\sum_{\eta=1}^{N}\sum_{\eta'\neq \eta}^{N}\left(\me^{-\mi \vec{k}_{0}\cdot \vec{r}_{\eta}}\me^{\mi \vec{k}_{0}\cdot \vec{r}_{\eta'}}+\me^{\mi \vec{k}_{0}\cdot \vec{r}_{\eta}}\me^{-\mi \vec{k}_{0}\cdot \vec{r}_{\eta'}}\right),\\
    &\langle \hat{\boldsymbol{Y}}^{2}_{1\beta} \rangle=\frac{(N-1)}{2}+\frac{1}{2(d-1)}+\frac{1}{2(d-1)N}\sum_{\eta=1}^{N}\sum_{\eta'\neq \eta}^{N}\left(\me^{-\mi \vec{k}_{0}\cdot \vec{r}_{\eta}}\me^{\mi \vec{k}_{0}\cdot \vec{r}_{\eta'}}+\me^{\mi \vec{k}_{0}\cdot \vec{r}_{\eta}}\me^{-\mi \vec{k}_{0}\cdot \vec{r}_{\eta'}}\right).
\end{align}
For $\alpha\neq 1<\beta$ and $\beta\geq 3$, we obtain the simple relations:
\begin{align}
    &\langle \hat{\boldsymbol{X}}^{2}_{\alpha\beta} \rangle=\frac{1}{(d-1)},\\
    &\langle \hat{\boldsymbol{Y}}^{2}_{\alpha\beta} \rangle=\frac{1}{(d-1)}.
\end{align}

Using the previous quantities, $w_{1,\vec{k}_{0}}$ takes the form:

\begin{align}
     \nonumber&1-\left(\frac{1}{d-1}\right)+2(d-1)\left(\frac{(N-1)}{2}+\frac{1}{2(d-1)}+\frac{1}{(d-1)N}\left(\left|\sum_{\eta=1}^{N}e^{\mi \vec{k}_{0}\cdot \vec{r}_{\eta}}\right|^{2}-N\right)\right)\\
&+\left(\frac{d(d-1)}{2}-(d-1)\right)\left(\frac{2}{(d-1)}-\frac{2}{(d-1)^{2}N^{2}}\left|\sum_{\eta=1}^{N}e^{\mi \vec{k}_{0}\cdot \vec{r}_{\eta}}\right|^{2}\right)\geq (d-1)N.
\end{align}
Invoking the structure factor notation (i.e, $S_{\vec{k}_{0}}$), it is equivalent to
\begin{align}
    -\frac{2(N-1)S_{\vec{k}_{0}}+d(N^{2}+S_{\vec{k}_{0}}-2NS_{\vec{k}_{0}})}{(d-1)N^{2}}\geq 0.
\end{align}
From the last expression, we get the condition for entanglement
$S_{\vec{k}_{0}}\leq N^{2}/(2N-1)$. Interestingly, this condition is independent of the internal dimension of the atom $d$.
\section{Light polarization dependent detection}

Let us consider the quantum state
\begin{align}
    \ket{\psi}=\frac{1}{\sqrt{3}}(\ket{11}+\ket{22}+\ket{13}).
\end{align}
It follows that the linear operators satisfy:
\begin{align}
    &\langle \hat{\boldsymbol{X}}_{12} \rangle=0,\\
    &\langle \hat{\boldsymbol{X}}_{13} \rangle=\frac{1}{3\sqrt{2}|\zeta_{13}|}\left(e^{-\mi \vec{k}_{0}\cdot \vec{r}_{2}}\zeta_{13}+e^{\mi \vec{k}_{0}\cdot \vec{r}_{2}}\zeta_{31}\right),\\
    &\langle \hat{\boldsymbol{X}}_{23} \rangle=0.
\end{align}
\begin{align}
    &\langle \hat{\boldsymbol{Y}}_{12} \rangle=0,\\
    &\langle \hat{\boldsymbol{Y}}_{13} \rangle=\frac{\mi}{3\sqrt{2}|\zeta_{13}|}\left(e^{-\mi \vec{k}_{0}\cdot \vec{r}_{2}}\zeta_{13}-e^{\mi \vec{k}_{0}\cdot \vec{r}_{2}}\zeta_{31}\right),\\
    &\langle \hat{\boldsymbol{Y}}_{23} \rangle=0.
\end{align}

\begin{align}
    &\langle \hat{\boldsymbol{Z}}_{11} \rangle=1,\ \ \langle \hat{\boldsymbol{Z}}_{22} \rangle=\frac{2}{3},\ \ \langle \hat{\boldsymbol{Z}}_{33} \rangle=\frac{1}{3}.
\end{align}
In a similar way, we get for higher-order momenta:
\begin{align}
    &\langle \hat{\boldsymbol{X}}^{2}_{12} \rangle=\frac{5}{6}+\frac{1}{3|\zeta_{12}|^{2}}\left(e^{-\mi \vec{k}_{0}\cdot \vec{r}_{1}}e^{-\mi \vec{k}_{0}\cdot \vec{r}_{2}}\zeta^{2}_{12}+e^{\mi \vec{k}_{0}\cdot \vec{r}_{1}}e^{\mi \vec{k}_{0}\cdot \vec{r}_{2}}\zeta^{2}_{21}\right),\\
    &\langle \hat{\boldsymbol{X}}^{2}_{13} \rangle=\frac{4}{6},\\
    &\langle \hat{\boldsymbol{X}}^{2}_{23} \rangle=\frac{1}{2}.
\end{align}
\begin{align}
    &\langle \hat{\boldsymbol{Y}}^{2}_{12} \rangle=\frac{5}{6}-\frac{1}{3|\zeta_{12}|^{2}}\left(e^{-\mi \vec{k}_{0}\cdot \vec{r}_{1}}e^{-\mi \vec{k}_{0}\cdot \vec{r}_{2}}\zeta^{2}_{12}+e^{\mi \vec{k}_{0}\cdot \vec{r}_{1}}e^{\mi \vec{k}_{0}\cdot \vec{r}_{2}}\zeta^{2}_{21}\right),\\
    &\langle \hat{\boldsymbol{Y}}^{2}_{13} \rangle=\frac{4}{6},\\
    &\langle \hat{\boldsymbol{Y}}^{2}_{23} \rangle=\frac{1}{2}.
\end{align}
\begin{align}
    &\langle \hat{\boldsymbol{Z}}^{2}_{11} \rangle=1+\frac{2}{3},\ \ \langle \hat{\boldsymbol{Z}}^{2}_{22} \rangle=\frac{2}{3}+\frac{2}{3},\ \ \langle \hat{\boldsymbol{Z}}^{2}_{33} \rangle=\frac{1}{3}.
\end{align}

Now we compute the modified second-order operators and variance:

\begin{align}
    &\langle \bar{\boldsymbol{X}}^{2}_{12} \rangle=\frac{1}{3|\zeta_{12}|^{2}}\left(e^{-\mi \vec{k}_{0}\cdot \vec{r}_{1}}e^{-\mi \vec{k}_{0}\cdot \vec{r}_{2}}\zeta^{2}_{12}+e^{\mi \vec{k}_{0}\cdot \vec{r}_{1}}e^{\mi \vec{k}_{0}\cdot \vec{r}_{2}}\zeta^{2}_{21}\right),\\
    &\langle \bar{\boldsymbol{X}}^{2}_{13} \rangle=0,\\
    &\langle \bar{\boldsymbol{X}}^{2}_{23} \rangle=0.
\end{align}
\begin{align}
    &\langle \bar{\boldsymbol{Y}}^{2}_{12} \rangle=-\frac{1}{3|\zeta_{12}|^{2}}\left(e^{-\mi \vec{k}_{0}\cdot \vec{r}_{1}}e^{-\mi \vec{k}_{0}\cdot \vec{r}_{2}}\zeta^{2}_{12}+e^{\mi \vec{k}_{0}\cdot \vec{r}_{1}}e^{\mi \vec{k}_{0}\cdot \vec{r}_{2}}\zeta^{2}_{21}\right),\\
    &\langle \bar{\boldsymbol{Y}}^{2}_{13} \rangle=0,\\
    &\langle \bar{\boldsymbol{Y}}^{2}_{23} \rangle=0.
\end{align}
\begin{align}
    &\langle \bar{\boldsymbol{Z}}^{2}_{11} \rangle=\frac{2}{3},\ \ \langle \bar{\boldsymbol{Z}}^{2}_{22} \rangle=\frac{2}{3},\ \ \langle \bar{\boldsymbol{Z}}^{2}_{33} \rangle=0.
\end{align}
\begin{align}
    & (\Delta \bar{\boldsymbol{Y}}_{12})^{2}=-\frac{1}{3|\zeta_{12}|^{2}}\left(e^{-\mi \vec{k}_{0}\cdot \vec{r}_{1}}e^{-\mi \vec{k}_{0}\cdot \vec{r}_{2}}\zeta^{2}_{12}+e^{\mi \vec{k}_{0}\cdot \vec{r}_{1}}e^{\mi \vec{k}_{0}\cdot \vec{r}_{2}}\zeta^{2}_{21}\right),\\
    &(\Delta \bar{\boldsymbol{Y}}_{13})^{2}=-\left(\frac{\mi}{3\sqrt{2}|\zeta_{13}|}\left(e^{-\mi \vec{k}_{0}\cdot \vec{r}_{2}}\zeta_{13}-e^{\mi \vec{k}_{0}\cdot \vec{r}_{2}}\zeta_{31}\right)\right)^{2},\\
    &(\Delta \bar{\boldsymbol{Y}}_{23})^{2}=0.
\end{align}
Using a specific case of the second inequality given by Eq.\eqref{eq:app_second} (with $N=2$), we get
\begin{align}
   \nonumber &\sum_{\mu_{-}}(\Delta \bar{\boldsymbol{Y}}_{\mu_{-}})^{2} - \sum_{\mu_{z}}\langle \bar{\boldsymbol{Z}}^{2}_{\mu_{z}}\rangle -\sum_{\mu_{+}}\langle \bar{\boldsymbol{X}}^{2}_{\mu_{+}}\rangle \geq -2.
\end{align}
with $1\leq \mu_{-},\mu_{+}\leq 3$ and $1 \leq \mu_{z}\leq 3$ we obtain
\begin{align}
    &\nonumber-\frac{1}{3|\zeta_{12}|^{2}}\left(e^{-\mi \vec{k}_{0}\cdot \vec{r}_{1}}e^{-\mi \vec{k}_{0}\cdot \vec{r}_{2}}\zeta^{2}_{12}+e^{\mi \vec{k}_{0}\cdot \vec{r}_{1}}e^{\mi \vec{k}_{0}\cdot \vec{r}_{2}}\zeta^{2}_{21}\right)+\frac{1}{18|\zeta_{13}|^{2}}\left(e^{-\mi \vec{k}_{0}\cdot \vec{r}_{2}}e^{-\mi \vec{k}_{0}\cdot \vec{r}_{2}}\zeta^{2}_{13}+e^{\mi \vec{k}_{0}\cdot \vec{r}_{2}}e^{\mi \vec{k}_{0}\cdot \vec{r}_{2}}\zeta^{2}_{31}-2|\zeta_{13}|^{2}\right)\\
    &-\frac{4}{3}-\frac{1}{3|\zeta_{12}|^{2}}\left(e^{-\mi \vec{k}_{0}\cdot \vec{r}_{1}}e^{-\mi \vec{k}_{0}\cdot \vec{r}_{2}}\zeta^{2}_{12}+e^{\mi \vec{k}_{0}\cdot \vec{r}_{1}}e^{\mi \vec{k}_{0}\cdot \vec{r}_{2}}\zeta^{2}_{21}\right)\geq -2.
\end{align}
The difference between the left-hand and right-hand quantities is plotted in Fig.\ref{fig:1} for different configurations.

\section{Remark 3: On the mirror symmetry between measured right and left circular polarizations}

\begin{proof}
    Let us consider that there are $n$ allowed transitions with a defined dipole moment $\vec{d}_{\alpha,\beta}$ $\forall (\alpha,\beta) \in n$. For each transition, we define a polarization factor ${\zeta_{1,+}},\hdots,\zeta_{n,+}$ (when all light coming from the $n$  transitions is measured along the right-circular polarization component  $\hat{{e}}_{+}$) and $\zeta_{1,-},\hdots,\zeta_{n,-}$ (when all light coming from the  $n$ transitions is measured along the left-circular polarization component $\hat{{e}}_{-}$).  We know that the only polarization-dependent inequalities are $w^{\bar{\boldsymbol{X}},\bar{\boldsymbol{Y}},\bar{\boldsymbol{Z}}}_{2,\vec{k}_{0}}$, $w^{\bar{\boldsymbol{Y}},\bar{\boldsymbol{X}},\bar{\boldsymbol{Z}}}_{2,\vec{k}_{0}}$, $w^{\bar{\boldsymbol{X}},\bar{\boldsymbol{Z}},\bar{\boldsymbol{Y}}}_{3,\vec{k}_{0}}$ and $w^{\bar{\boldsymbol{Y}},\bar{\boldsymbol{Z}},\bar{\boldsymbol{X}}}_{3,\vec{k}_{0}}$. An important property of the previous inequalities, is that all of them can be defined as a general function $f\in \mathbb{R}$  in the following way

\begin{align}
   &F_{1} = f((\zeta_{1,+})^{2},\hdots,(\zeta_{n,+})^{2},(\zeta^{*}_{1,+})^{2},\hdots,(\zeta^{*}_{n,+})^{2},|\zeta_{1,+}|^{2},\hdots,|\zeta_{n,+}|^{2})\geq 0,\\
   &F_{2}=f((\zeta_{1,-})^{2},\hdots,(\zeta_{n,-})^{2},(\zeta^{*}_{1,-})^{2},\hdots,(\zeta^{*}_{n,-})^{2},|\zeta_{1,-}|^{1},\hdots,|\zeta_{n,-}|^{2})\geq 0.
\end{align}
Considering the case where $\vec{d}_{\alpha\beta}\in \mathbb{R}$ $\forall (\alpha,\beta)$, which implies that $\zeta_{\gamma,+}=-(\zeta_{\gamma,-})^{*}$ and $|\zeta_{\gamma,+}|^{2}=|\zeta_{\gamma,-}|^{2}$ $\forall \gamma \in \{1,2,\hdots,n\}$. Now, taking an arbitrary complex number $\zeta_{\gamma,+}=r_{\gamma}\me^{\mi \phi_{\gamma}}$ (i.e. $\zeta_{\gamma,-}=-r_{\gamma}\me^{-\mi \phi_{\gamma}}$), we get

\begin{align}
   &F_{1} = f(r^{2}_{1}\me^{2\mi \phi_{1}},\hdots,r^{2}_{n}\me^{2\mi \phi_{n}},r^{2}_{1}\me^{-2\mi \phi_{1}},\hdots,r^{2}_{n}\me^{-2\mi \phi_{n}},|\zeta_{1,+}|^{2},\hdots,|\zeta_{n,+}|^{2})\geq 0,\\
   &F_{2}=f(r^{2}_{1}\me^{-2\mi \phi_{1}},\hdots,r^{2}_{n}\me^{-2\mi \phi_{n}},r^{2}_{1}\me^{2\mi \phi_{1}},\hdots,r^{2}_{n}\me^{2\mi \phi_{n}},|\zeta_{1,+}|^{1},\hdots,|\zeta_{n,+}|^{2})\geq 0.
\end{align}
From the last expressions, we conclude that
\begin{align}
    F_{1}(\phi_{1},\hdots,\phi_{n})=F_{2}(-\phi_{1},\hdots,-\phi_{n}).
\end{align}
which proves our Remark \ref{remark:3}. This result also implies that there will be a symmetry between the pairs of measured polarizations $(\hat{{e}}_{+},\hat{{e}}_{z})$ and $(\hat{{e}}_{-},\hat{{e}}_{z})$, since $\hat{{e}}_{z}$ introduces only a real parameter.
\end{proof}

\end{document}